\documentclass[10pt,journal,compsoc]{IEEEtran}
\usepackage{multirow}

\ifCLASSOPTIONcompsoc
  \usepackage[nocompress]{cite}
\else
  \usepackage{cite}
\fi
\ifCLASSINFOpdf
   \usepackage[pdftex]{graphicx}
\else
   \usepackage[dvips]{graphicx}
\fi
\usepackage{amsmath}
\usepackage[ruled,vlined]{algorithm2e}
\usepackage{algorithmic}

\usepackage{array}

\ifCLASSOPTIONcompsoc
 \usepackage[caption=false,font=footnotesize,labelfont=sf,textfont=sf]{subfig}
\else
 \usepackage[caption=false,font=footnotesize]{subfig}
\fi

\usepackage{color}

\ifCLASSOPTIONcaptionsoff
 \usepackage[nomarkers]{endfloat}
\let\MYoriglatexcaption\caption
\renewcommand{\caption}[2][\relax]{\MYoriglatexcaption[#2]{#2}}
\fi
\usepackage{url}

\newcommand{\sky}[1]{\textcolor{black}{#1}}
\newcommand{\skyR}[1]{\textcolor{black}{#1}}

\begin{document}

\title{FAAG: Fast Adversarial Audio Generation through Interactive Attack Optimisation}
%
%
%
%

\author{Yuantian Miao\IEEEauthorrefmark{1},
Chao Chen\IEEEauthorrefmark{2},
Lei Pan\IEEEauthorrefmark{3},
Jun Zhang\IEEEauthorrefmark{1}, 
Yang Xiang\IEEEauthorrefmark{1}
\IEEEcompsocitemizethanks{\IEEEcompsocthanksitem\IEEEauthorrefmark{1}School of Software and Electric Engineering, Swinburne University of Technology, Hawthorn, VIC 3122, Australia\hfil\break
\IEEEcompsocthanksitem\IEEEauthorrefmark{2}Corresponding Author, College of Science and Engineering, James Cook University, Townsville, QLD 4811, Australia\hfil\break
\IEEEcompsocthanksitem\IEEEauthorrefmark{3}School of Information Technology, Deakin University, Geelong, VIC 3220, Australia\hfil\break
}}

\IEEEtitleabstractindextext{%
\begin{abstract}
Automatic Speech Recognition services (ASRs) inherit deep neural networks' vulnerabilities like crafted adversarial examples. Existing methods often suffer from low efficiency because the target phases are added to the entire audio sample, resulting in high demand for computational resources. This paper proposes a novel scheme named FAAG as an iterative optimization-based method to generate targeted adversarial examples \skyR{quickly}. By injecting the noise over the beginning part of the audio, FAAG generates adversarial audio in high quality with a high success rate timely. Specifically, we use audio's logits output to map each character in the transcription to an approximate position of the audio's frame. \sky{Thus, an adversarial example can be generated by FAAG in approximately two minutes using CPUs only and around ten seconds with one GPU while maintaining an average success rate over 85\%.} \skyR{Specifically, the FAAG method can speed up around 60\% compared with the baseline method during the adversarial example generation process.} Furthermore, we found that appending benign audio to any suspicious examples can effectively defend against the targeted adversarial attack. We hope that this work paves the way for inventing new adversarial attacks against speech recognition with computational constraints.
\end{abstract}

}

\maketitle

\IEEEdisplaynontitleabstractindextext

%

\section{Introduction}\label{sec:introduction}
\IEEEPARstart{A}{utomatic} speech recognition (ASR) technologies have enabled the transformation of human spoken language into text. In recent years, with the development of advanced deep learning techniques, the efficiency and effectiveness of ASR systems are enhanced as a Deep-Learning-as-a-Service. ASR service has become an increasingly popular human-machine interface due to its accuracy and efficiency/convenience. The number of devices with voice assistants is estimated to reach 8.4 billion by 2024 from the current 4.2 billion globally \cite{chuck_2020}. The value of the global ASR market will be over USD 21.5 billion by 2024 \cite{chuck_2020}. International corporate giants like Microsoft, Google, IBM, and Amazon, are heavily investing in new technologies to expand their market shares. Devices and applications integrating the acoustic systems are ubiquitous, like Amazon Echo, Google Assistant, and Apple's Siri, enabling the full potential of intelligent voice-controlled devices, voice personal assistants, and machine translation services \cite{du2019sirenattack, abdullah2020faults}. Hence, the security problems associated with ASR systems are worth millions of dollars. 

With the advancement of deep neural networks, ASR systems have become increasingly prevalent in our daily lives  \cite{lokesh2018automatic,mehrabani2015personalized,schonherr2018adversarial,liu2019adversarial}. Despite ASR's popularity, the security risk and the adversarial attack against ASRs have raised concerns in the security community \cite{zhang2019dangerous,tung2019exploiting,kumar2018skill,wang2020security,liu2018detecting}. The community has confirmed that ASR systems inherit vulnerabilities from neural networks \cite{lin2020software}. For example, neural network models are vulnerable to adversarial examples \cite{biggio2013evasion,bruna2013intriguing}. In addition, state-of-the-art ASR systems consisting of deep neural network structures can be fooled by adversarial examples \cite{cisse2017houdini}. Machine learning-based cybersecurity has become an important challenge in various real-world applications \cite{coulter2020data,sun2018data,chen2020android,qiu2020survey}. 

Existing research shows that well-crafted adversarial audio can lead an ASR system to misbehave unexpectedly. There are two types of adversarial attacks --- targeted attacks and untargeted attacks. Untargeted attacks against ASR systems can damage the performance of the ASR system. Abdullah et a\sky{l}.~\cite{abdullah2019hear} forced an ASR system to transcribe the input audio into incorrect texts. Targeted attacks against ASR systems not only cause low accuracy in transcription but also inject the attacker's desired phrases without being recognized. Carlini et al.~\cite{carlini2016hidden} leveraged noise-like hidden voice commands to embed commands into a normal audio example so that users can only hear a meaningless white noise, but the ASR system can execute the hidden commands. The DolphinAttack further crafted the audio to make the embedded command inaudible and imperceptible to human beings. Carlini and Wagner \cite{carlini2018audio} proposed an interactive optimization-based method to enable an adversarial example generated in a small distortion of Decibels ($dB$). Qin et al.~\cite{qin2019imperceptible} improved the method using the psychoacoustic principle of auditory masking to generate unnoticeable noise.

Although various methods have been proposed to generate adversarial audio in high quality, these methods may not perform as well as expected under certain conditions. Firstly, \sky{all previous methods assume abundant time to generate the adversarial audio in high quality with multiple GPUs. However, when provided with limited resources, e.g., only CPUs or just one available GPU, previous attacks are almost always more time-consuming than expected. Especially for a bunch of adversarial examples, a fast adversarial audio generation method could be more dangerous than the existing ones.} Secondly, all existing methods use a complete audio clip to generate the adversarial example. However, as users' security awareness has gradually increased nowadays, an adversarial audio clip may not be played entirely if any anomaly is noticeable. When this audio was only played partly, the success rate of the attack would be decreased significantly because the command is not fully loaded. Thus, an adversarial audio generation method based on a target audio segment is more powerful than the one based on the complete audio. The shorter the target audio segment, the more robust the attack.

This paper aims to find an effective and efficient method for adversarial attack under white-box access to the target end-to-end ASR system. Our method aims to improve the existing popular method in \cite{carlini2018audio} based on previous concerns. We propose to modify a part of an audio example, instead of its whole frame. The beginning part of the audio is large enough to be covered by the target phrase to embed any phrases into an audio example, including voice commands. Only a space separating the target phrase and the remaining transcription texts are needed to ensure the ASR system can understand the target phrase. Thus, our method's key task is to find a proper length of the frame at the proper position of the audio. The state-of-the-art ASR systems can filter out some noise and rectify some contextual errors recognized by the system using their language model. Therefore, it is difficult to hide any phrases correctly as a part of original audio's transcription. A fixed length at the beginning part of the audio is the best solution.
Otherwise, there is not enough space to embed the target phrase and result in a low success rate. Specifically, the proper clip of the audio frame is selected by mapping each word in the transcription to a rough position in the audio's frame according to the logits output. This audio clip is subsequently used to generate the adversarial example based on the interactive optimization-based method proposed by \cite{carlini2018audio}. The contributions of this work can be summarized as follows:
\begin{itemize}
    \item We propose a new scheme, called Fast Adversarial Audio Generation (FAAG), developing a new optimisation algorithm based on an interactive attack strategy. According to different phrases and provided audios, FAAG can automatically select a proper length of the frame at the beginning of the audio. The shortest ratio of the frame used for an adversarial example generation can be reached to 14.79\%. 
    \item We develop a fast adversarial sample generation with a satisfied success rate and tolerable distortion. Under a limited resources, our method takes half an hour using the CPUs only to generate ten adversarial examples, and cause at around two minutes using one GPU. Both are \sky{faster} than previous attacks using the same resources\skyR{, specifically speeding up around 60\% in generation time}.
    \item The empirical study provides us with two new observations: (1) different words in target phrases will not significantly affect the performance of our adversarial examples; (2) a target phrase with fewer words has a slight positive boost on the adversarial example generation, compared to a target phrase containing more words.
    \item The target phrase can only be hidden at the beginning of the original audio, otherwise the transcription of the phrase part will be at low accuracy. On the contrary, appending a begin audio at the beginning of any suspicious audio can effectively protect the service from the targeted adversarial attack.
\end{itemize}

The rest of the paper is organized as follows: Section~\ref{sec:bg} introduces the background about the target ASR model and related adversarial attacks against the ASR system. Section~\ref{sec:method} illustrates the details of our method to generate audio adversarial examples with limited resources. Section~\ref{sec:exp} presents the setup and the results of our experiments. Section~\ref{sec:discussion} discuss different positions of audio using our method and countermeasures. Finally, Section~\ref{sec:conclude} concludes the paper.

\section{Related Work}\label{sec:bg}
This section introduces the state-of-the-art Automatic Speech Recognition (ASR) models and the related work about adversarial attacks on ASRs.

\subsection{The Automatic Speech Recognition Model}
While conventional ASR models are based on hidden Markov models (HMMs), current state-of-the-art ASR models utilise deep neural networks (DNNs). Our audio adversarial attack targets a state-of-the-art ASR system based on a DNN --- end-to-end ASR systems~\cite{perero2018exploring}. Assuming white-box access, we evaluate our attack using an ASR model downloaded from a popular open-source ASR system DeepSpeech.

\begin{figure}[t] 
\begin{center}
\centerline{\includegraphics[width=0.5\textwidth]{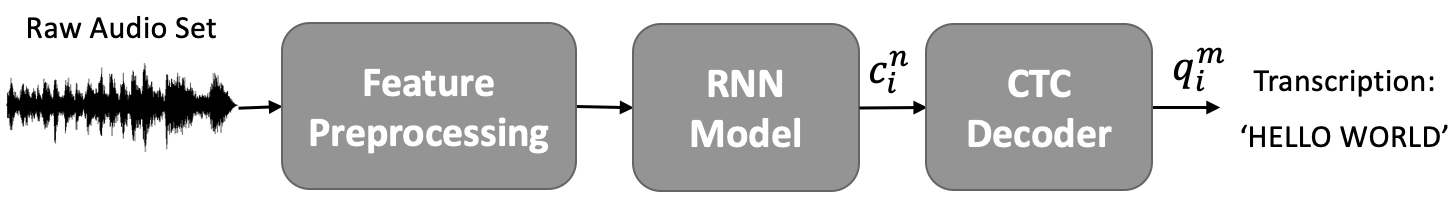}}

\caption{An end-to-end ASR system. There are three main components: (i) the pre-processing step extracts MFC features to represent the raw audio data and is separated into a few window frames, (ii) an RNN is usually used to transcribe each audio frame at a time as one character with its probability distributions, (iii) the CTC decoder decodes the characters output $c_{i}^{N}$ and rescores it to a word sequence $q_{i}^{M}$, since the CTC allows unknown alignment between the input and output sequences.}
\label{fig:E2E_system}
\end{center}
\end{figure}

\textbf{End-to-end ASR systems} in Baidu's DeepSpeech implemented by Mozilla are sequence-to-sequence neural network models~\cite{hannun2014deep, amodei2016deep}. Unlike other typical hybrid ASR systems, the end-to-end system predicts word sequences that are converted directly from individual characters from the raw waveform. As shown in Fig.~\ref{fig:E2E_system}, the end-to-end system is a unified neural network modeling framework containing three main components, including a feature pre-processing step, a neural network model as the probabilistic model, and a decoder to refine the final outputs. 

The feature pre-processing step uses Mel-Frequency Cepstral Coefficients (MFCC) features to represent the raw audio data. The method in the first step is Mel-Frequency Cepstrum (MFC) transformation. The whole frame with the MFCC features extracted will be split into multiple frames with an overlapping window. Each audio frame will be fed into the probabilistic model. Herein, Recurrent Neural Networks (RNNs) are popular in End-to-End ASR systems where an audio waveform is mapped to a sequence of characters \cite{hannun2014deep}. However, the sequence of the character output $c_{i}^{N}$ does not mean the sequence of the word output $q_{i}^{M}$. Thus, the decoder is used to reevaluate the character output. Connectionist Temporal Classification (CTC) \cite{graves2006connectionist} is a powerful method for the unknown alignment between the input and output sequence. DeepSpeech uses CTC as a decoder to score the character output and map it to a word sequence by de-duplicating sequentially repeated characters. 

\begin{figure*}[t] 
\begin{center}
\centerline{\includegraphics[width=0.8\textwidth]{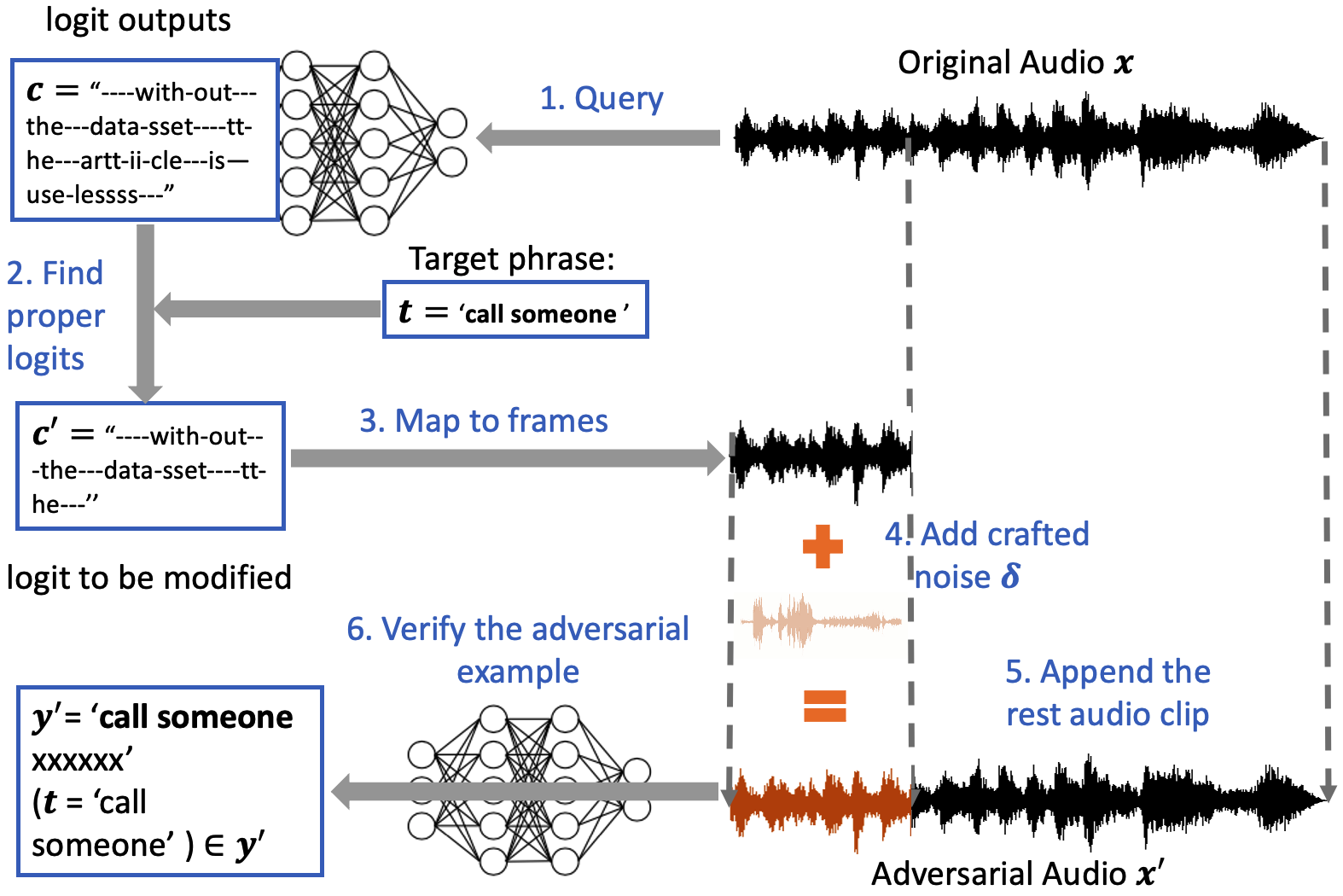}}
\vskip 0.1in
\caption{The overall process of \sky{Fast Adversarial Audio Attack (FAAG).}
1) Given an audio $x$, query the ASR model with white-box access and gain its logit outputs. 2) Find the proper logits whose number of words restrict the number of words in the target phrase. 3) According to the window size defined in ASR system's preprocessing step, we map the proper logits to their corresponding frames in the original audio as $x_{begin}$. 4) Add crafted noise $\delta$ to $x_{begin}$ and let $x'_{begin} = x_{begin} + \delta$ be translated as the target phrase $t$. 5) Appending the rest audio clip $x_{rest}$ to the modified clip $x'(begin)$, we gain the adversarial audio  $x'=x'_{begin}+x_{rest}$ whose transcription result contains our target phrase. 6) The adversarial example is validated by querying the adversarial example to the ASR model. The attack is considered successful if $t \in y'$.}
\label{fig:attack_overall}
\end{center}
\end{figure*}

\subsection{Adversarial Attack on ASRs}
Audio adversarial attacks on ASR systems employ both targeted and untargeted adversarial attacks \cite{abdullah2020faults}. The knowledge of these attacks' target model is set at white-box access or black-box access. Generally, the audio adversarial attack aims to generate an audio adversarial example to deceive the ASR model without users' awareness. When the ASR model simply mistranslates an audio example, it is an untargeted adversarial attack. When the ASR model translates an audio example into a phrase designed by the attacker, it is a targeted adversarial attack. 

Unlike the image domain, the targeted adversarial attack is much more dangerous against ASR systems than the untargeted adversarial attack. Except a few examples like an untargeted adversarial attack on an ASR system to force mistranscription in \cite{abdullah2019hear}, most attacks focus on targeted adversarial attacks, where the target phrase is usually a common voice command \cite{carlini2016hidden}.

Targeted adversarial attacks with white-box access can generate adversarial examples of high quality. Specifically, \cite{carlini2018audio} generated an adversarial audio example with only slight distortion on the \textit{DeepSpeech} model. CommanderSong \cite{yuan2018commandersong} can embed the desired voice commands into any songs stealthily. \textit{SirenAttack} is proposed to generate adversarial audios under white-box and black-box settings \cite{du2019sirenattack}. Under the white-box setting, \textit{SirenAttack} applies a fooling gradient method to find the adversarial noise, whose success rate can reach to 100\%. \sky{The perturbed MFCC features can be reversed to attack ASR into adversarial speech \cite{xiao2018generating}.} Different from \cite{xiao2018generating}, Gong et al.~\cite{gong2017crafting} generated adversarial audios by modifying the raw waveform directly with an end-to-end scheme. Furthermore, the adversarial examples could be generated with different systems and different features according to \cite{kreuk2018fooling}. Audio adversarial examples are designed in \cite{qin2019imperceptible} by leveraging the psychoacoustic principle of auditory masking.

\sky{Taking an End-to-End ASR model --- \textit{DeepSpeech} model as the target model, Carlini and Wagner \cite{carlini2018audio} proposed an iterative optimization-based attack. In essence, a small noise was added over the complete benign audio and iteratively modified the noise via minimizing the loss function. An audio clip and the target phrase were fed as inputs into the target model. The loss function was set to CTC loss. To obtain imperceptible adversarial audio, a distortion metric was considered during the minimization process to ensure minimal noise. Further, the loss function was improved by fine-tuning the noise based on the alignments' probabilities. The highest success rate was reportedly 100\%. Based on this method, Qin et al.~\cite{qin2019imperceptible} added auditory masking to the noise, considering the psychoacoustic principle for deriving imperceptible adversarial audio. Yakura and Sakuma \cite{akura2019robust} implemented Carlini and Wagner's method in the over-the-air condition by modifying the environmental reverberation in a physical setup.}

Targeted adversarial attacks with black-box access are more practical than white-box attacks. With little knowledge of the ASR system, Hidden Voice Commands generates the noisy command by repeatedly querying the model \cite{carlini2016hidden}. As a result, the semantics of the generated adversarial audio is difficult for people to understand. \textit{DolphinAttack} exploits the non-linearity of the microphones to generate inaudible voice commands \cite{zhang2017dolphinattack}. Under the black-box attack, \textit{SirenAttack} proposed an iterative and gradient-free method \cite{du2019sirenattack}. The works in \cite{taori2019targeted, alzantot2018did} considered genetic algorithms and gradient estimation to modify the original audio under black-box access. The work in \cite{schonherr2018adversarial} generated adversarial examples based on psychoacoustic hiding in the black-box access, which can embed any audio with a malicious voice command. Devil's whisper \cite{chen2020devil} proposed a general adversarial attack against the ASR systems by training a local model under white-box access.


Different from the related work, this paper focuses on the efficiency of adversarial speech recognition. We propose to modify a part of an audio example through interactive attack optimization to guarantee a high success rate, low distortion, and high generation speed.

\section{Generating Audio Adversarial Examples} \label{sec:method}
\subsection{Threat Model}
\sky{As mentioned above, state-of-the-art audio adversarial attacks can already generate a high-quality adversarial audio clip with a high success rate. However, two limitations are neglected. Firstly, all previous attacks assume the attacker has multiple GPUs and abundant time to generate such high-quality adversarial audio. Once the computing sources are not that intensive, the time spent on attack would increase significantly. However, a successful audio adversarial attack requires timely action in the real world. Secondly, as the user's awareness of security and privacy is enhanced, even a slight noise within the audio would be noticed. In such cases, the shorter the adversarial audio can be recognized as the target phrase, the more powerful the adversarial audio is. We propose the FAAG method to hide the target phrase in a small piece of adversarial audio quickly with limited computing sources supported.}

Given an audio waveform $x$ and the target transcription $t$, we aim to construct an adversarial audio $x' = x + \delta$. Assuming that the target ASR system transcribes both the audio $x$ into text $y$ ($y=\mathcal{F}(x)$) and the audio $x'$ as text $y'$ ($y'=\mathcal{F}(\sky{x'})$), we expect that the target transcription $t$ is a sub-string of $y'$. Additionally, the audio $x$ and our adversarial audio $x'$ should sound normal to human beings. We formulate the similarity based on the difference of the distortion in Decibels (dB) between the original audio $x$ and the crafted adversarial audio $x'$, which is represented by the noise $\delta$, the same as \cite{carlini2018audio}. The $dB$ value is formulated in Section~\ref{sec:exp}. The $dB$ difference ($dB_{x}(\delta)$) between the modified adversarial example and the original audio reflects the relative loudness of the added noise comparing to the original audio. The smaller their difference in $dB$ is, the more similar that two audio examples sound. When the loudness of noise is small enough, the noise can be ignored so that the adversarial audio can be transcribed by the ASR system without human awareness of the attack. 

We assume that the adversarial audio is generated with white-box access to the target ASR model. Herein, the attacker has complete knowledge of the ASR model, including its structure and its parameters. In our paper, we choose to use Baidu's \textit{DeepSpeech} model \cite{hannun2014deep} that is a popular and accessible ASR model open-sourced by Mozilla. DeepSpeech includes three parts --- an MFC conversion for audio preprocessing, RNN layers to map each input frame into a probability distribution over each character, and a CTC loss function to measure the RNN's output score. The core of the \textit{DeepSpeech} model is an optimized RNN trained by multiple GPUs on over 5,000 hours of speech from 9,600 speakers. The RNN layers finally output logits that are computed over the probability distribution of output characters. 

We do not consider over-the-air audio transcription because many live settings may jeopardize the experiments. \sky{We will discuss our method adapted in the real world in Section~\ref{sec:discussion}.} FAAG only modifies a small portion of the given audio example instead of the whole frame.
It will be challenging to compare FAAG with other methods using live transcription. Furthermore, our adversarial examples are validated by transcribing the waveform directly. We treat the adversarial attack as a successful attack if the output transcription $y'$ includes the target phrase $t$ correctly, denoted by $t \in y'$.

\subsection{Fast Adversarial Audio Generation (FAAG)}
\begin{algorithm}[t]
\caption{Select the Proper Clip $x_{begin}$}
\label{alg:select_audio}
\begin{algorithmic}
\REQUIRE Original audio $\boldsymbol{x}$; Target phrase $\boldsymbol{t}$; Pre-trained ASR model $\mathcal{F}$; Step = s; Fine-tune variable $\lambda$
\STATE $\boldsymbol{y} = \mathcal{F}(\boldsymbol{x})$
\STATE $\boldsymbol{c} = f(\boldsymbol{x})$
\STATE $\boldsymbol{t}$ = $\boldsymbol{t}$ + `  '
\STATE $|\dots|$ presents the character number or the frame length in the vector $\dots$.
\ENSURE The selected audio clip $\boldsymbol{x_{begin}}$ is long enough for adversarial example generation.
\STATE Initialize $\lambda = 0$
\STATE $|\boldsymbol{t_{allocated}}| = |\boldsymbol{t}| + \lambda$
\STATE $|\boldsymbol{x_{begin}}| \geq \frac{|\boldsymbol{c}|}{|\boldsymbol{y}|} \times |\boldsymbol{t_{allocated}}| \times s$
\STATE $index = |\boldsymbol{x_{begin}}|$
\STATE $\boldsymbol{x_{begin}} \gets \boldsymbol{x}[:index]$
\STATE $\boldsymbol{x_{rest}} \gets \boldsymbol{x}[index:]$
\RETURN Two clips $x_{begin}$ and $x_{rest}$
\end{algorithmic}
\end{algorithm}

\sky{Fast Adversarial Audio Generation (FAAG) is an intelligent and efficient adversarial audio example generation method. FAAG generates an adversarial audio clip within a short period, even with limited computing sources, while embedding the attacker's desired command (the target phrase) in a short clip of this adversarial audio. To shorten the adversarial audio clip within the whole audio, we try to find a proper position and length of frames used to embed our target phrase with a high success rate and low distortion. }We find that it is unnecessary to transcribe the whole audio waveform as our selected phrase. \sky{When generating an adversarial audio example, the longer the waveform used does not lead to less distortion or higher success rate.} However, the longer the waveform that is used, the slower the adversarial example will be generated. By constructing the targeted phrase at the beginning of the adversarial example and separating the targeted phrase from the rest transcription with a long space, the ASR model can still recognize the targeted phrase correctly. Comparing to the prior work on targeted attacks on speech-to-text \cite{carlini2018audio}, we generate the audio adversarial example based on the beginning part of a long audio waveform instead of the whole waveform.

Fig.~\ref{fig:attack_overall} illustrates our adversarial example generation, focusing on a targeted adversarial attack on the DeepSpeech model. In general, given an audio waveform $x$, the target ASR model's transcription is $y$ and our adversarial attack can be summarized as three steps. 

Step 1: Based on any chosen short phrase $t$, we select the proper frames at the beginning of $x$ as $x_{begin}$ to add noise $\delta$. \sky{We choose the beginning of the audio because no prior noise would affect the accuracy, and the effect of the subsequent clip's noise could be limited.} Herein, we describe the phrase $t$ as a short phrase when its length is shorter than the length of the given transcription $y$, denoted by $|\boldsymbol{t}| < |\boldsymbol{y}|$. 

Step 2: We construct the inaudible noise $\delta$ with an iterative and optimization-based attack. Therefore, $x_{begin} + \delta$ can be recognized by the ASR model as the phrase $t$ with a specific conjunction (i.e.~ `and') or a long space. Herein, the long space means a silence recognized by the ASR model. It is necessary so that the transcription $y'$ of the adversarial example excluding the phrase $t$ will not affect the model's understanding of our chosen phrase $t$. 

Step 3: We combine $x'_{begin} = x_{begin} + \delta$ with the rest of the frames of $x$ (named as $x_{rest} = x - x_{begin}$) so that the adversarial example $x' = x'_{begin} + x_{rest}$ sounds similar to the original audio $x$. In addition, the adversarial example $x'$ is recognized as $y'$ by the ASR model where $t \in y'$. To evaluate the success rate of the adversarial example, we calculate the character error rate (CER) of $t$ in $y'$.

We explain the three steps in the following subsections.

\subsubsection{Selecting the proper frames at the beginning of a given audio.} 
We define the proper frames $x_{begin}$ to satisfy three conditions: 1) the frames used to generate adversarial examples should be at the beginning of the original audio $x$; 2) the length of the frames should be long enough to cover the target phrase $t$ correctly; 3) the generated adversarial examples should have relatively small distortion. Thus, we choose the frames of audio $x$ corresponding to the first $n$ words of its transcription $y$, where the number of words in the target phrase is $n = len(t)$. To meet the second condition, we consider the frame length of each logit and the amount of logits for $n$ words. Therefore, we need to find out the relationship between the input audio $x$, its logit output $c$ and its corresponding transcription $y$. As for the third condition, we add a variable $\lambda$ to fine tune the length to result in a small distortion.

As for the first condition, generating the adversarial examples as the beginning parts of the audio has some advantages. Firstly, as the beginning parts of the audio, the target phrase can be recognized by the ASR model with less effect from the remaining original audio frames than if it were inserted in the middle of the audio. Secondly, it is easier for the ASR model to recognize and execute the target phrase, especially when the target phrase is a command. For example, when the victim plays our crafted adversarial audio, and the ASR model recognizes a command hidden at the beginning of the audio, it is very likely to execute the command regardless of the remaining audio's meaning. However, there are some special cases. For example, when the target phrase contains a trigger word of an ASR system, the ASR system will only listen to the sentence behind that trigger word. The position of the adversarial example hidden in the original audio becomes less important than in previous models. Thus, we consider the case of generating an adversarial audio clip at the middle and at the end of the original audio in later experiments. To shed light on the method of our attack, we take the beginning position as an example.

\begin{figure}[t] 
\begin{center}
\centerline{\includegraphics[width=0.5\textwidth]{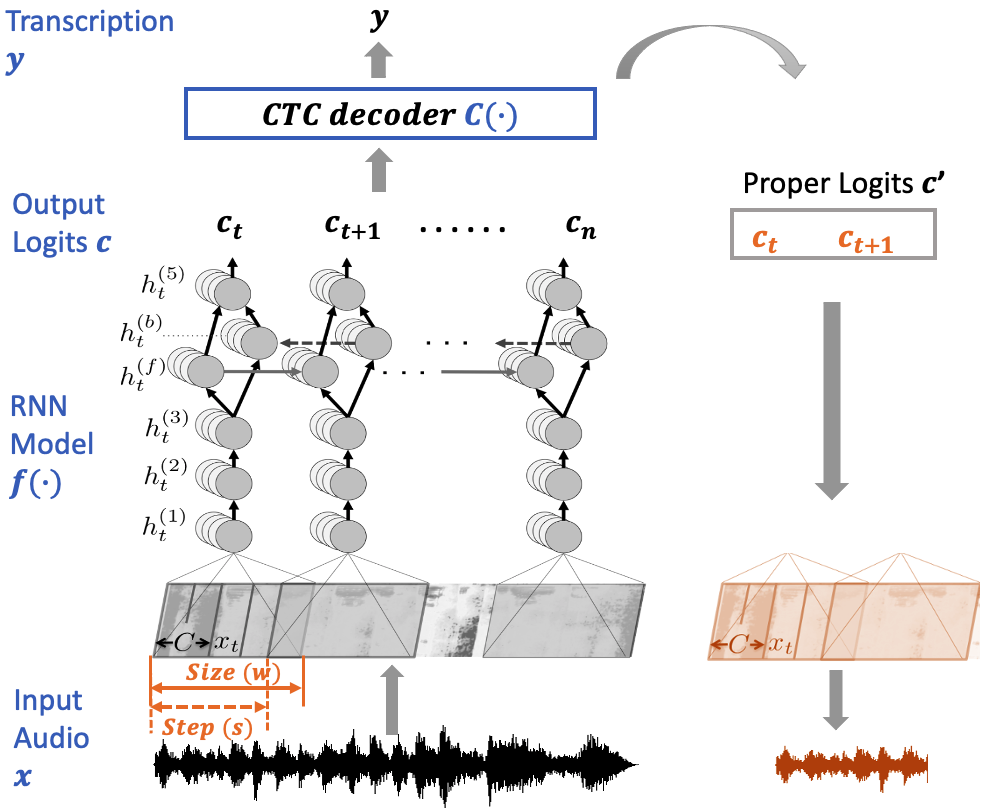}}
\vskip 0.1in
\caption{The \textit{DeepSpeech} model transcribing an input audio $x$ as its transcription $y$. \sky{The proper frame length is selected based on the logit outputs.} 
}
\label{fig:frames_logits}
\end{center}
\end{figure}

To satisfy the second condition, we need clarify the relationship between $x$, $c$, and $y$. Thus, it is necessary to understand the mechanism of the target ASR model. The target ASR model in this work is Baidu's DeepSpeech model \cite{hannun2014deep}, specifically an end-to-end Speech-to-Text model implemented by Mozilla. Fig.~\ref{fig:frames_logits} demonstrates the diagram of the DeepSpeech model transcribing an input audio $x$ as its transcription $y$. The input audio $x$ firstly is split from a whole frame into several overlapping windows with a window size $w$. Herein, each window slips to the next window with a step length of $s$. After the MFC transformation, the RNN model $f(\cdot)$ in DeepSpeech maps each output logit $c_{i} \in \boldsymbol{c}$ as a probability distribution over each character during each window frame ($\|\boldsymbol{c}\|=\|\boldsymbol{x}\|$). The character $c_{i}$ is in range `a' to `z', white space, and the $'-'$ symbol which represents the epsilon value $\epsilon$ in CTC decoding. Then the CTC decoder $C(\cdot)$ outputs a sequence of characters $\boldsymbol{y}$ with an overall probability distribution, merges repeats and drops epsilons. To decode a vector $\boldsymbol{c}$ to a transcription vector $\boldsymbol{y}$, the best alignment can be found by Equation~\ref{eq:ctc} according to \cite{carlini2018audio}.
\begin{equation} \label{eq:ctc}
    C(\boldsymbol{x}) = argmax_{\boldsymbol{y}}Pr(\boldsymbol{y}\|f(\boldsymbol{x}))
\end{equation}

We can summarize the relationship between $\boldsymbol{x}$, $\boldsymbol{c}$, and $\boldsymbol{y}$ in line with the mechanism of the target ASR model. The whole frame of input audio $x$ is spilt into several overlapping windows with a step of length $s$. The RNN model $f(\cdot)$ in DeepSpeech maps each output logit $c_{i}$ as a probability distribution over each character within each window frame. Then, we can conclude the relationship between the output logits $\boldsymbol{c}$ and the whole frame of the input audio $\boldsymbol{x}$ as per Equation~\ref{eq:c&x}. Herein, the $|\cdot|$ represents the number of characters within a text (i.e. $|\boldsymbol{c}|$) or the length of frames in the audio (i.e.~$|\boldsymbol{x}|$).
\begin{equation} \label{eq:c&x}
    |\boldsymbol{x}| = |\boldsymbol{c}| \times s + (|\boldsymbol{x}| \mod s)
\end{equation}
Naturally, the more characters the transcription has, the longer the length of the output logits. There is a positive correlation between between the output logits $\boldsymbol{c}$ and transcription $\boldsymbol{y}$. We summarise this as Equation~\ref{eq:c&y}.
\begin{equation} \label{eq:c&y}
    \rho(|\boldsymbol{c}|, |\boldsymbol{y}|) > 0
\end{equation}

Translating audios with the same ASR model, we assume that the relationship between our generated adversarial audio clip $\boldsymbol{x'_{begin}}$, the corresponding output logits $\boldsymbol{c'}$, and the corresponding transcription $\boldsymbol{y'}$ is the same as that between $\boldsymbol{x}$, $\boldsymbol{c}$, and $\boldsymbol{y}$. According to Equation~\ref{eq:c&y}, $\rho(|\boldsymbol{c'}|, |\boldsymbol{y'}|) > 0$. However, the transcription $y'$ is different from the original transcription $y$. Between the output logits and the transcription, the CTC decoder $C(\cdot)$ will merge repeats and drop epsilons within the logits to get the final transcription. The number of repeats and epsilons within the logits varies in lines with the speaker's speaking habits and the ASR model's window size. Although the window size is the same, the speaker's speaking habits are hard to control in different recordings. We assume that $\rho(|\boldsymbol{c'}|, |\boldsymbol{y'}|) \approx \rho(|\boldsymbol{c}|, |\boldsymbol{y}|)$. To simplify the experiment, we refine this relationship into the following equations.
\begin{align}
  \frac{|\boldsymbol{c}|}{|\boldsymbol{y}|} &= \frac{|\boldsymbol{c'}|}{|\boldsymbol{y'}|} \label{eq:c&y2} \\
  |\boldsymbol{x'_{begin}}| &= |\boldsymbol{c'}| \times s + (|\boldsymbol{x'_{begin}}| \mod s) \\
  &= \frac{|\boldsymbol{c}|}{|\boldsymbol{y}|} \times |\boldsymbol{y'}| \times s + (|\boldsymbol{x'_{begin}}| \mod s) \label{eq:x'}
\end{align}
Analogous to Equation~\ref{eq:c&x}, we know the relationship between frame length of $x'_{begin}$ and the number of output logits. Combining this with Equation~\ref{eq:c&y2}, we can find the frame length of $x'_{begin}$  as per Equation~\ref{eq:x'}. Assuming our adversarial example generation is successful, the transcription of our adversarial audio clip $y'$ is the same as our target phrase $t$. The frame length of the selected audio clip should be the same as the frame length of our generated adversarial audio clip ($|\boldsymbol{x_{begin}}| = |\boldsymbol{x'_{begin}}|$). Thus, we can find a proper minimum length of the frames for adversarial audio generation from Equation~\ref{eq:x'} to meet the second condition. At least, we know the range of proper frame length selected from the beginning of the original audio (as shown in Equation~\ref{eq:x'_range}). 
\begin{equation} \label{eq:x'_range}
   \frac{|\boldsymbol{c}|}{|\boldsymbol{y}|} \times |\boldsymbol{t}| \times s + s > |\boldsymbol{x_{begin}}| = |\boldsymbol{x'_{begin}}| \geq \frac{|\boldsymbol{c}|}{|\boldsymbol{y}|} \times |\boldsymbol{t}| \times s
\end{equation}
We need to ensure that the selected length is long enough for adversarial example generation. In this work, we set  ($|\boldsymbol{x_{begin}}| = \frac{|\boldsymbol{c}|}{|\boldsymbol{y}|} \times |\boldsymbol{t}| \times s$) during experiments and use a variable $\lambda$ to fine tune the frame length of the selected audio clip. Using Algorithm~\ref{alg:select_audio}, the original audio can be split into two audio clips including $\boldsymbol{x_{begin}} = \boldsymbol{x}\left[: |\boldsymbol{x_{begin}}|\right]$ and $\boldsymbol{x_{rest}} = \boldsymbol{x}\left[|\boldsymbol{x_{begin}}|:\right]$. With the proper frame length selected, not only the time is saved, but also the negative effect of the remaining audio on the adversarial example's transcription can be neglected. Accordingly, the success rate of the generated adversarial example can be increased.

Apart from the guarantee of the success rate, we also consider selecting a proper frame length to satisfy the third condition --- less distortion in the adversarial audio. Normally, using the same generation method based on an audio clip with a fixed frame length, the more characters the target phrase has, the more distortion occurs in the generated adversarial audio because with more characters in the target phrase, more characters in the original audio are needed to be changed. However, targeting a phrase based on a different length of audio, the longer audio may not produce adversarial audio with less distortion. We introduce a new variable $\lambda$ to fine tune the proper frame length at the beginning of the original audio and discuss it in the next section.

\begin{algorithm}[t]
\caption{Audio Adversarial Example Generation}
\label{alg:adversarial_generation}
\begin{algorithmic}
\REQUIRE Original audio $x$; Target transcription $t$; Pre-trained model $\mathcal{F}$; WS = $s$; $iter=1,000$
\ENSURE $len(y) \geq len(t)$
\STATE Call Algorithm ~\ref{alg:select_audio}
\RETURN Two audio clips: $x'_{begin}$ and $x_{rest}$
\STATE $dB(x_{begin}) = 20*log_{10}(np.max(np.abs(x_{begin})))$
\STATE Optimize $\delta$: $con$ is a constant to narrow down the dB
\STATE $dB_{x_{begin}}(\delta) = dB(\delta) - dB(x_{begin})$
\FOR{iteration range from 1 to $iter$}
\WHILE{$\mathcal{F}(\delta) \neq t$ and $dB_{x'_{begin}}(\delta) > con$}
\STATE $x_{begin} \gets x_{begin} - w \cdot sign(\Delta_{x_{begin}}ctc\_L(x_{begin}, t))$
\STATE $x'_{begin} \gets x_{begin}$
\STATE minimize $dB_{x_{begin}}(\delta) + \sum_{i}w_{i} \cdot ctc\_L(x'_{begin}))$
\ENDWHILE
\IF{$\mathcal{F}(x'_{begin}) == t$ and $dB_{x_{begin}}(\delta) \leq con$ and $iter/100 = 0$}
\STATE $con \leftarrow con \times 0.8$
\ENDIF
\ENDFOR
\STATE 
\STATE $x' = x'_{begin} + x_{rest}$
\STATE $time = end\_time - start\_time$
\STATE Calculate its distortion: $dB_{x}(\delta) = dB(x') -  dB(x)$
\STATE Verify: $y' \leftarrow \mathcal{F}(x')$
\STATE Success if $t \in y'$
\RETURN Adversarial audio $x'$; Adversarial transcription $y'$; distortion in Decibels $dB_{x}(\delta)$; time
\end{algorithmic}
\end{algorithm}

\subsubsection{Constructing inaudible noise in the proper frames} 
Knowing the proper audio segment $x_{begin}$, we construct inaudible noise $\delta$ and generate $x'_{begin} = x_{begin} + \delta$. Based on the optimization method proposed in \cite{carlini2018audio}, we optimize $\delta$ according to the CTC loss function $ctc\_L(\cdot)$ with a constraint on $dB_{x}(\delta)$ mentioned below. The optimization method can be summarized as Equation \ref{eq:optimize}. Herein, $w_{i}$ represents the relative importance of being close to $t$ and remaining close to $x_{begin}$. $c_{i}$ is a character of the output logits processed by the RNN model, while the $dB_{x}(\delta)$ indicates the difference of dB between the original audio and the noise. The constant $con$ is initially a large constant value, and will be reduced to run the minimization again till the result converges. Finally, the output of this step is constructed as $x'_{begin}$.
\begin{equation} 
\begin{split}
\label{eq:optimize}
    \text{minimize } & \|\delta\|_{2}^{2} + \sum_{i} w_{i} \cdot ctc\_L(x_{begin} + \delta, c_{i})\\
    \text{such that } & dB_{x}(\delta) \leq con
\end{split}
\end{equation}

According to \cite{carlini2018audio}, the minimization problem is solved using an Adam optimization with 100 learning rate and 1,000 iterations. As shown in Algorithm~\ref{alg:adversarial_generation}, $x_{begin}$ will be updated with the CTC loss function as $x'_{begin}$ varies. For every 100 iterations, if the current adversarial example is successful ($\mathcal{F}(x'_{begin} == t)$ \& $dB_{x_{begin}}(\delta) \leq con$), we narrow down the constant $con$ by $con = con \times 0.8$ to search for even smaller distortions. Different from the distortion calculated in \cite{carlini2018audio}, we combine the modified $x'_{begin}$ with the remaining audio $x_{rest}$ and subsequently calculate the $dB$ of the whole adversarial example marked as $x'=x'_{begin}+x_{rest}$ before obtaining the distortion as $dB_{x}(\delta) = dB(x') - dB(x)$.

\subsubsection{Adversarial example generation and evaluation} 
With $x'_{begin}$ generated, the adversarial example is generated by combining this clip with $x_{rest}$. Thus, the final adversarial example is $x' = x'_{begin} + x_{rest}$. The whole process is defined in Algorithm~\ref{alg:adversarial_generation}. We define success by verifying the transcription recognized by the target ASR model with an adversarial example. Specifically, we consider the successfully generated adversarial example $x'$, when its transcription $y'$ contains the short phrase $t'$ and exactly matches the target phrase ($t = t'$). When $t \neq t'$, we say that our adversarial example is generated without 100\% accuracy. In this case, we measure its success rate with character error rate (CER) defined in Section~\ref{sec:exp}.

\section{Evaluation}\label{sec:exp}
\subsection{Experimental Setting}
We set up a number of experiments to evaluate our proposed adversarial example generation. Ten audios are selected randomly from the TIMIT dataset as the target audio to generate the adversarial example, which can be recognized by a pre-trained DeepSpeech model as our target phrase. At the same time, the change is inaudible to human beings. All the experiments are conducted on a workstation with an Intel Core X i9-7960X CPU (16 Cores) and 128GB memory. Some experiments use one TITAN XP GPU in addition to the CPUs.

\subsubsection{Dataset Description}
The \textbf{TIMIT} speech corpus (TIMIT) is a famous speech corpus to build and evaluate ASR systems. Specifically, 630 speakers across the United States recorded audios for this corpus, including 6,300 sentences \cite{garofolo1993timit}. Each speech waveform in this corpus is sampled at 16-bit, 16kHz for each utterance. In this work, we propose an effective method of injecting the target phrase into a long waveform. Thus, we randomly select five audios waves whose transcriptions have relatively more words than our target phrases. Herein, we define the audio's transcription with more than ten words as relatively more words. This setting applied to our target phrases is because the number of words in most sensitive commands is less than ten. For example, the command ``call someone'' and ``turn on the airplane mode'' have a handful of words. Hence, it is reasonable to select these audio waveforms as our target audios.

\subsubsection{Target ASR Model}
The \textbf{DeepSpeech} ASR model is our target, which is an end-to-end ASR model with the CTC loss function applied \cite{hannun2014deep, amodei2016deep}. Herein, RNN is the core engine to translate audio to a sequence of text. Specifically, the pre-trained model we targeted is \texttt{deepspeech-0.4.1-model}. This speech-to-text model is trained with multiple corpora, including \textit{LibriSpeech}, \textit{Fisher}, \textit{Switchboard}, and \textit{English common voice training corpus}. It can reach 8.26\% WER when testing on the \textit{LibriSpeech} dataset. The training batch size is 24; the testing batch size is 48; the learning rate is 0.0001; the dropout rate is 0.15; and the number of neurons in the hidden layer is 2,048.

\subsubsection{Baseline}
Since our adversarial example generation is an improvement based on Carlini and Wagner's work \cite{carlini2018audio}, we generate adversarial examples using the iterative optimization-based method proposed by \cite{carlini2018audio} as the baseline. According to the implementation in \cite{carlini2018audio}, any audio may be translated into any phrase. \cite{carlini2018audio} applied the perturbations over the complete frames of the original audio. They solved the optimization problem using the Adam optimizer with a learning rate of 10. The default iteration to generate an audio adversarial example in this work is 1,000, while the maximum iteration in \cite{carlini2018audio} is 5,000. They can generate targeted adversarial examples with 100\% success rate with a mean distortion from $-31dB$ to $-38dB$. Our generated adversarial example only modifies the beginning of the original audio, but the baseline method modifies the original audio frames. \skyR{Both FAAG and the baseline method are evaluated on the same workstation for a fair comparison.}

\subsubsection{Target Phrases}
Apart from comparing our generation method with the baseline, we also evaluate FAAG's effectiveness in modifying audio results in different target phrases. Specifically, we define two sets of target phrases where each set contains three phrases listed in Table~\ref{tab:target_phrase}. The first set is used to evaluate our generation method's performance by injecting different words of the target phrases containing three words each into the original audio. We name this set as the \textit{target phrases with different words}. Another set named as the \textit{target phrases with different lengths} is used to evaluate our generation method's performance by injecting target phrases of different lengths into the original audio.

\begin{table}[t]
\caption{Two sets of target phrases to evaluate the audio adversarial generation.}
\label{tab:target_phrase}
\centering
\begin{tabular}{|c|l|c|}
\hline
\textbf{Two Sets} & \multicolumn{1}{c|}{\textbf{Target Phrase}} & \multicolumn{1}{l|}{\textbf{\# of Characters}} \\ \hline
\multirow{3}{*}{\begin{tabular}[c]{@{}c@{}}Target Phrases \\ with \\ Different Words\end{tabular}} & call john smith & 15 \\ \cline{2-3} 
 & call david jone & 15 \\ \cline{2-3} 
 & play music list & 15 \\ \hline
\multirow{3}{*}{\begin{tabular}[c]{@{}c@{}}Target Phrases\\ with\\ Different Lengths\end{tabular}} & call john smith & 15 \\ \cline{2-3} 
 & call john & 10 \\ \cline{2-3} 
 & call john smith and david & 25 \\ \hline
\end{tabular}
\end{table}

\subsubsection{Evaluation Metrics}
We evaluate our adversarial example generation method from three aspects, including the attack's success rate, the dB level of the noise $\delta$ compared to the original audio $x$, and the time needed under the limited resource. The three metrics are described in detail as follows:
\begin{itemize}
    \item The success rate of injecting the target phrase $t$ into the modified adversarial example can be calculated with the character error rate (success rate = 1 - $CER$). Assuming that the predicted target phrase is $t'$, the $CER$ is the ratio of the number of incorrect characters predicted in $t'$ over the total number of characters in $t$.
    \item The distortion in Decibels ($dB$) is used to quantify the distortion of the modified adversarial example comparing to the original one. According to \cite{carlini2018audio}, each audio's $dB$ is their relative loudness represented in a logarithmic way as $dB(x) = \max_{i} 20 \times \log_{10} (x_{i})$. In addition, the difference of $dB$ between the original audio and the noise can be formulated as $dB_{x}(\delta) = dB(x') - dB(x)$ \cite{carlini2018audio}. It is always hard to determine whether the modified audio is imperceptible to human beings using $dB$. We provide a benchmark of an adversarial example's distortion using method in \cite{carlini2018audio}. According to \cite{noise_2019}, $dB \approx 30dB$ is similar to the loudness of whisper.
    \item Considering the audio clip selected to generate the adversarial audio, ratio of frames is used to measure the ratio of the selected clip to the complete audio. The larger the ratio of frames, the longer the audio clip is clipped for adversarial example generation. When the ratio of frames is 100\%, the whole audio is used for generation. In this case, FAAG does not claim to select the best length of an audio clip for performing  adversarial attacks, consistent with the baseline methods \cite{carlini2018audio}.
    \item \skyR{Generation} time is another important metric to evaluate our generation method's effectiveness, especially when the resource is limited. In our work, we only employ CPUs and one GPU to generate audio adversarial examples. 
\end{itemize}

\subsection{Proper Frame Length Selection}
As we stated in previous section, a proper frame length selected for an adversarial audio clip generation should satisfy three conditions. The first condition is about the selected clip's position in the original audio. The second condition is related to the proper minimum number of windows mapping to our audio clip's frame length. The third condition adds a new variable $\lambda$ to fine tune to frame length for less distortion. We explore the specific impact of these three factors on our proper frame length selection for adversarial audio generation.

\subsubsection{Proper Minimum Frame Length for Adversarial Audio Clip}
We firstly evaluate our method towards the second condition. That is, whether the frame length we selected is long enough to cover the target phrase correctly. According to Fig.~\ref{fig:frames_logits}, each window of the original audio will be translated to one logit character. As discussed in Section~\ref{sec:method}, we know the proper minimum frame length ($|\boldsymbol{x'_{begin}}| = |\boldsymbol{c'}| \times s = \frac{|\boldsymbol{c}|}{|\boldsymbol{y}|} \times |\boldsymbol{t}| \times s$) from the Equation~\ref{eq:x'_range}. Thus, the length of the adversarial audio clip can be determined by the number of characters in the target phrase. Keep the target phrase unchanged, we change the required length for the target phrase to alter the number of output logits $|\boldsymbol{c'}|$. Thus, one less number of characters may means a few less number of windows. Herein, the smaller frame length is only a few windows' size smaller than our selected proper minimum frame length. To evaluate the proper minimum frame length selection, we compare it with the results of a smaller frame length of the selected audio clip. Then this reference frame length is \[|\boldsymbol{x''_{begin}}| = \frac{|\boldsymbol{c}|}{|\boldsymbol{y}|} \times (|\boldsymbol{t}| - 1) \times s. \]

In addition, to ensure the remaining phrase would not affect the translation of our adversarial audio clip, we add a word or a long space after the chosen phrase as our target phrase. Thus, along with different number of windows, we evaluate the necessity of a word or a long space being added to the chosen phrase. Herein, a long space means the number of space should be larger than one.

We run FAAG and compare the results in five different settings. The target model is \texttt{deepspeech-0.4.1-model}. One phrase $t$ is chosen as part of the transcription of our adversarial audio. A word `and' and two spaces are added after the chosen phrase separately as two target phrases, marked as $t_{and}$ and $t_{spaces}$. The first two settings are selecting an audio clip with proper minimum frame length $|\boldsymbol{x'_{begin}}|$, when the target phrases are $t_{and}$ and $t_{spaces}$ respectively. We mark these two settings as $x'_{begin} \rightarrow t_{and}$ and $x'_{begin} \rightarrow t_{spaces}$. The second two settings are selecting an audio clip with a smaller frame length $|\boldsymbol{x''_{begin}}|$, when the target phrases are $t_{and}$ and $t_{spaces}$ respectively. We mark these two settings as $x''_{begin} \rightarrow t_{and}$ and $x''_{begin} \rightarrow t_{spaces}$. The fifth setting selects an audio clip with the minimum frame length $|\boldsymbol{x'_{begin}}|$, when the target phrase is $t$ with only one space appended. We mark this as $x'_{begin} \rightarrow t$. Ten audios are selected randomly from the TIMIT dataset which has a different distribution from the model's training corpus. For each audio, we repeated the experiment ten times to report the average result. 

\begin{table}[t]
\caption{Performance of Adversarial Generation on A Target Phrase with Different Frame Lengths.}
\label{tab:proper_frameLen}
\centering
\begin{tabular}{|l|l|l|}
\hline
\textbf{Setting} & \textbf{Average} $\mathbf{dB_{\delta}}$ & \textbf{Average Accuracy (\%)} \\ \hline
$x'_{begin} \rightarrow t_{and}$ & 30.87 & 90.37  \\ \hline
$x''_{begin} \rightarrow t_{and}$ & 36.28 & 89.11  \\ \hline
$x'_{begin} \rightarrow t_{spaces}$ & 38.55 & 94.7 \\ \hline
$x''_{begin} \rightarrow t_{spaces}$ & 43.01 & 93.33 \\ \hline
$x'_{begin} \rightarrow t$ & 39.73 & 81.19 \\ \hline
\end{tabular}
\end{table}

Table~\ref{tab:proper_frameLen} shows the averaged $dB_{\delta}$ of generated audio examples and averaged accuracy of translating these adversarial audio examples with different experiment settings. Comparing the experiment $x'_{begin} \rightarrow t_{and}$ with $x''_{begin} \rightarrow t_{and}$, not only the accuracy is decreased but also the distortion is increasing because of a smaller frame length of the audio clip. The same conclusion can be obtained by comparing the experiment $x'_{begin} \rightarrow t_{spaces}$ with $x''_{begin} \rightarrow t_{spaces}$. The reason is that generating an adversarial example with the same target phrase, less length of frames needs more effort to alter less number of original characters to all characters in the target phrase. The proper minimum frame length can be calculated using Equation~\ref{eq:x'_range}. In addition, by comparing experiments $x'_{begin} \rightarrow t_{and}$ and $x'_{begin} \rightarrow t_{spaces}$ with $x'_{begin} \rightarrow t$, a word or a long space is required to be added to ensure the accuracy of adversarial examples. The reason is that the last word of the target phrase is easy to be influenced by the rest phrase of the original audio because of the nature of an ASR model. When the target phrase is $t_{and}$, the average distortion is less than the result when the target phrase is $t_{spaces}$. We infer that more modifications are required to alter a character to a space symbol than to another character. Meanwhile, with a long space between the target phrase and the rest phrase, the translation in the target phrase part will be less influenced.

\subsubsection{Proper Fine-tune Length for Adversarial Audio Clip}
As we find the proper minimum frame length, we evaluate our method towards the third condition. Apart from the accuracy, it is important to achieve a less distortion of generated adversarial audio example to hide the attacker's intention. To fine tune the length for adversarial audio clip, we explore the relationship between the allocated frame length of an audio clip $|\boldsymbol{x^{allo}_{begin}}|$ and the number of characters in the target phrase $|\boldsymbol{t}|$ in different audios. Here, we introduce a variable $\lambda$ to alter the allocated frame length. Based on Equation~\ref{eq:x'_range}, we say the allocated frame length can be calculated as \[|\boldsymbol{x^{allo}_{begin}}| = \frac{|\boldsymbol{c}|}{|\boldsymbol{y}|} \times (|\boldsymbol{t}| + \lambda) \times s.\] \noindent Since we use the right equation of Equation~\ref{eq:x'_range}, we only consider the positive value of $\lambda$. Because of the relationship we stated in Equation~\ref{eq:c&y2}, the $\lambda$ should satisfy the condition where $|\boldsymbol{t}| + \lambda \leq |\boldsymbol{y}|$.

To clarify the relationship, we use the $ratio\_frame$ to measure the ratio of allocated frame length for the adversarial audio generation to the whole frame length of this audio. To find the best $\lambda$, we randomly select one original audio and three different phrases appended with two spaces. To avoid the impact of different lengths in the target phrase, we use the set of target phrases with different words stated in Table~\ref{tab:target_phrase}.

\begin{figure}[th]
\centering
\subfloat[Success rate]{\includegraphics[width=0.48\textwidth]{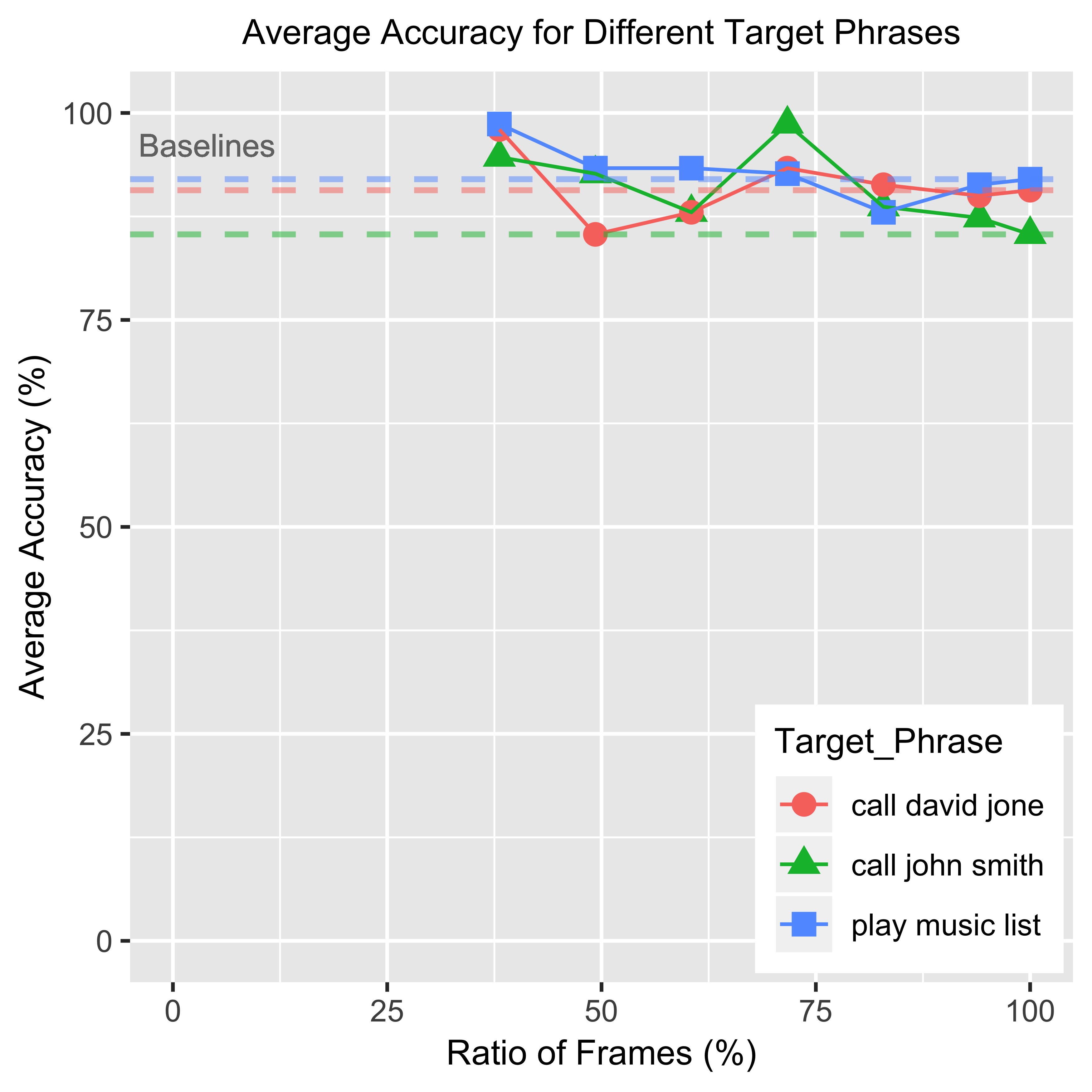}%
\label{subfig:frames_acc}}
\hfil
\subfloat[Distortion]{\includegraphics[width=0.48\textwidth]{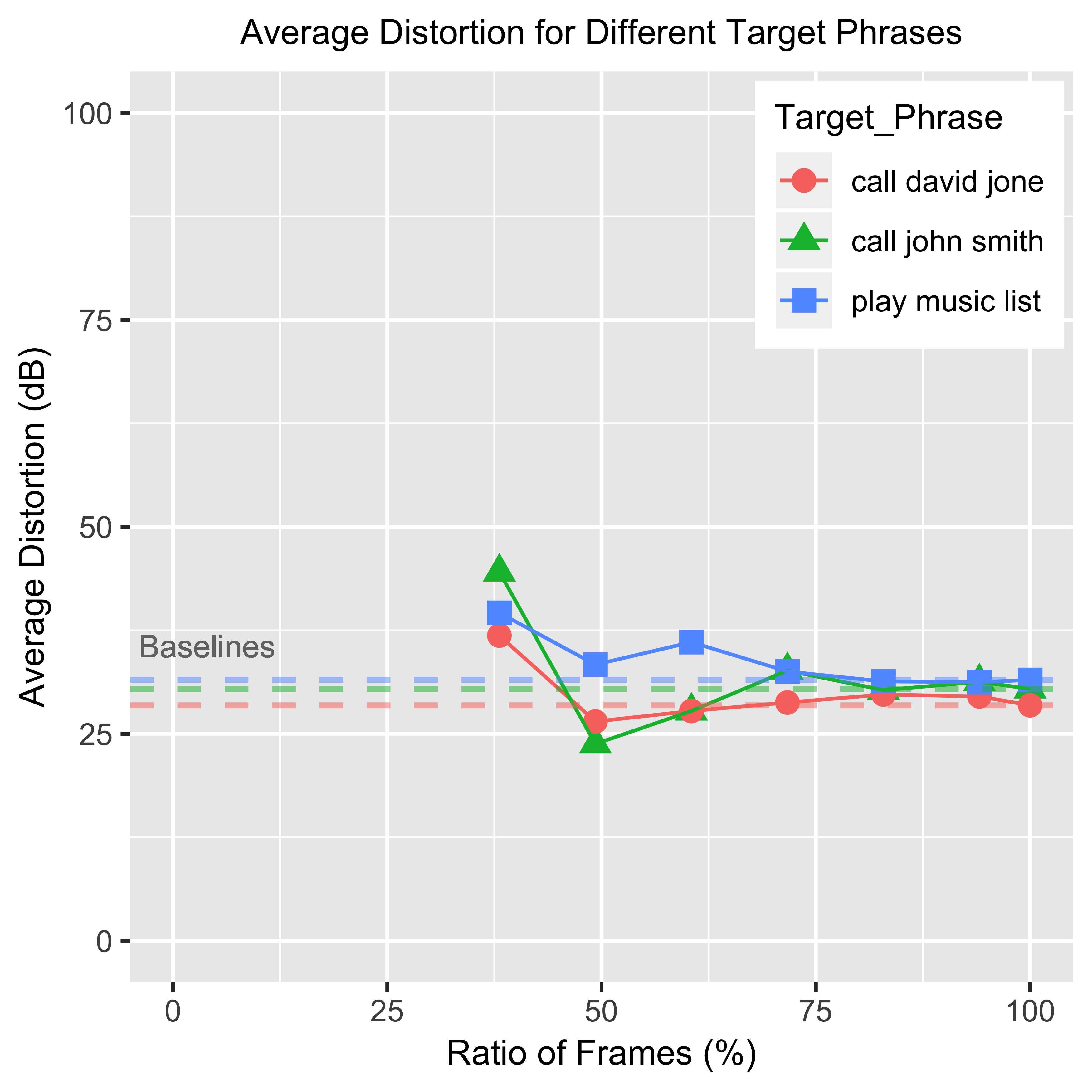}%
\label{subfig:frames_db}}
\caption{Performance of FAAG with the increasing length of selected audio clips for generation.}
\label{fig:finetune_perform}
\end{figure}

With the increasing ratio of frames clipped for adversarial audio generation, Fig.~\ref{fig:finetune_perform} shows the performance of our adversarial attack. When the ratio of the frame reaches to 100\%, the results are similar to our baseline. In general, the frame ratio can influence the success rate and distortion of generated adversarial audio. The best accuracy result outperforms by around 13\% than the baseline targeting the phrase `call john smith', while the best distortion is around 7dB smaller than the baseline. However, based on our observations, there are no specific rules about how $\lambda$ impacts the adversarial example's success rate and distortion. We infer that the reason may be the different efforts to transform various combinations of original characters to the target phrase. The greater the phoneme gap between the selected combination of original characters and the target phrase, the louder the noise is needed to generate our adversarial audio. In all, the best $\lambda$ to fine-tune the best length of a selected audio clip depends on the specific original audio and the target phrase.

\begin{figure}[th]
\centering
\subfloat[One original audio with three target phrases.]{\includegraphics[width=0.48\textwidth]{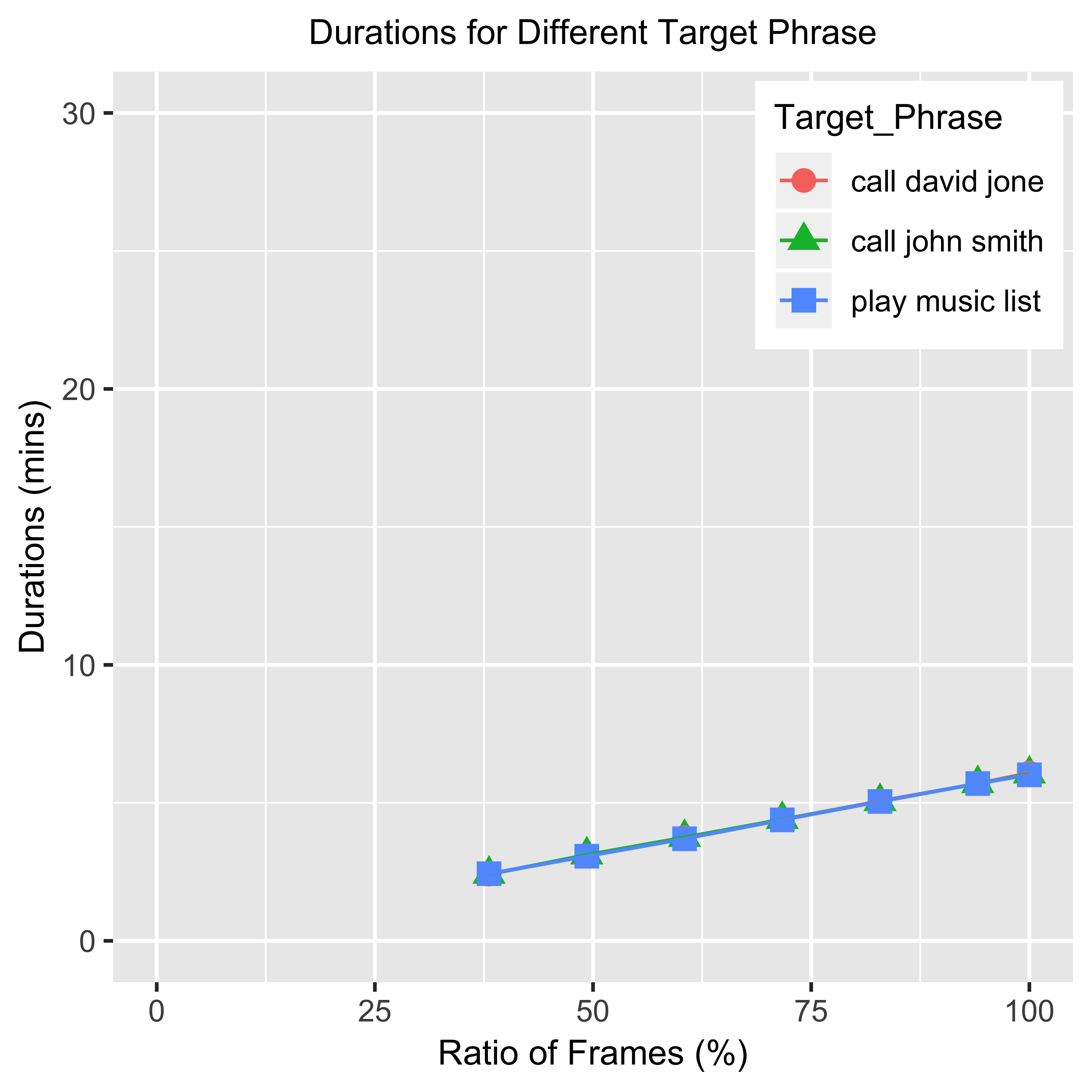}%
\label{fig:phrases_dur}}
\hfil
\subfloat[Different original audios with one target phrase.]{\includegraphics[width=0.48\textwidth]{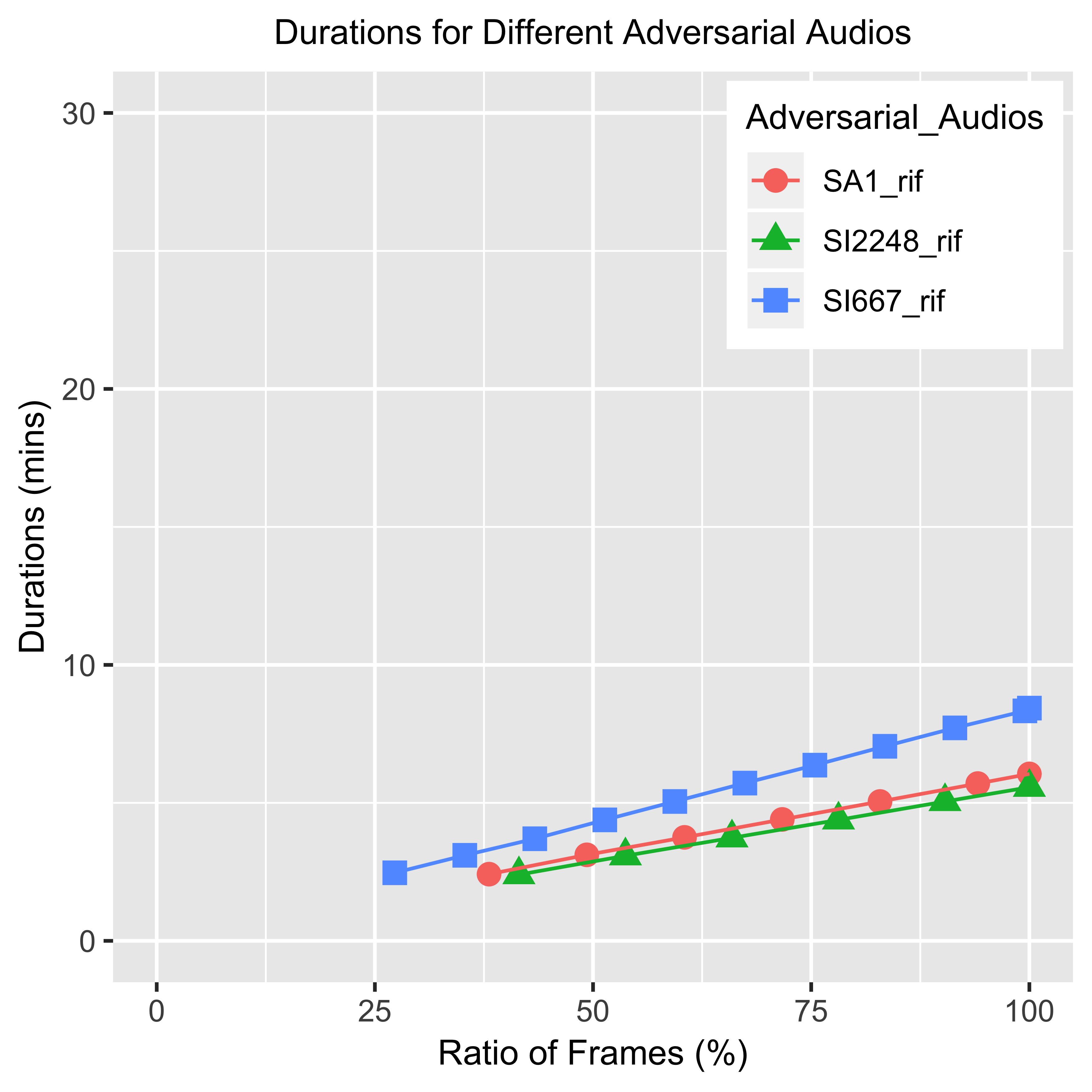}%
\label{fig:wavs_dur}}
\caption{Duration of FAAG generating each adversarial audio clip ten times.}
\label{fig:finetune_dur}
\end{figure}

Apart from the effective performance, we also consider the efficiency of our FAAG method. Apparently, from the results shown in Fig.~\ref{fig:finetune_dur}, we conclude that less time is spent using a shorter audio clip. To have further investigation, we generate adversarial examples using one original audio targeting three different phrases. Herein, these phrases are only different in words having the same length. As shown in Fig.~\ref{fig:phrases_dur}, different words in the target phrase will not affect the time spent on adversarial example generation. Additionally, we generate adversarial examples using three different original audios targeting the same phrase. Although different audios have different time requirements for FAAG, their growth rate in time is similar when the ratio of frames increases.

\subsection{Effectiveness and Efficiency Analysis}
We evaluate our generation method by comparing the generation performance of target phrases with different words, target phrases with different lengths, and comparing the generation performance with the baseline. Our experiments are conducted under a limited resource, as we only use CPUs and one GPU. In the rest results, when the ratio of frames is equal to 100\%, this experiment uses the baseline method to generate the adversarial example.

\subsubsection{Adversarial Generation of Target Phrases with Various Words}

\begin{table}[th]
\caption{Performance of Adversarial Generation on Target Phrases with Different Words. (Duration results are in Hour:Minute:Second format for all 100 adversarial audio generation. Apart from the duration, the other results are averaged.)}
\label{tab:generation_words}
\centering
\resizebox{.48\textwidth}{!}{
    \begin{tabular}{|p{0.1\textwidth}|p{0.1\textwidth}|p{0.05\textwidth}|p{0.07\textwidth}|p{0.08\textwidth}|}
    \hline
    \textbf{Target Phrase} & \textbf{Success Rate} & \textbf{$\mathbf{dB_{x}(\delta)}$} & \textbf{Duration} & \textbf{Ratio of Frames} \\ \hline
    \multirow{2}{*}{call john smith} & 89.07\% & 29 & 01:10:45 & 100\% \\ \cline{2-5} 
                                     & 91.33\% & 35 & 00:26:09 & 45.65\% \\ \hline
    \multirow{2}{*}{call david john} & 90.06\% & 27 & 01:11:08 & 100\% \\ \cline{2-5} 
                                     & 91.13\% & 32 & 00:26:29 & 45.65\% \\ \hline
    \multirow{2}{*}{play music list} & 90.39\% & 26 & 01:11:18 & 100\% \\ \cline{2-5} 
                                     & 90.67\% & 33 & 00:26:21 & 45.65\% \\ \hline
    \end{tabular}%
}
\end{table}

We generate our audio adversarial examples corresponding to the 100 TIMIT audio we selected randomly. To compare the adversarial attack performance on target phrases with different words, we choose the target phase the phrase set named target phrases with different in Table \ref{tab:target_phrase}. Moreover, we record their average results of success rate and distortion, while the duration time is counted for all 100 audio examples generation. Assuming that the attacker does not have enough time to generate the adversarial example, we only run the optimization-based method with 1,000 iterations to generate the adversarial example $x'_{begin} = x_{begin} + \delta$. Comparing with the baseline results, Table \ref{tab:generation_words} lists the performance of FAAG in the environment with limited computational resources and time. We analyze the results from the effectiveness and efficiency aspects.

For the effectiveness analysis, we focus on the success rate and distortion results in Table~\ref{tab:generation_words}. In general, updating the noise and distortion within 1,000 iterations, our FAAG method shares similar effectiveness as the baseline. All success rates are around or over 90\%, while the distortions are around 30. The impact of the different words in the target phrase can be ignored because of slight differences. As we discussed in the previous section, the audio clip's frame length will impact the whole adversarial example's performance. However, the impact relies on the specific original audio and target phrase. Attackers can fine-tune this factor using $\lambda$. In all, FAAG did not bring a negative impact on the effectiveness of the attack proposed in previous work. Different words in the target phrase do not have a particularly noticeable impact.

For the efficiency analysis, we focus on the duration and $ratio\_frame$. Generally, the ratio of frames used to generate the adversarial examples positively correlates to the duration required in the generation. Specifically, when the target phrase is `call john smith', almost half (45.65\%) of the original audio (100\%) is clipped for FAAG. About half of the time is spent on the FAAG compared to around one hour used in the baseline. When the target phrases are different in words, but the average ratio of frames is the same in the same length. Accordingly, the duration of FAAG targeting these three phrases is similar to each other. 

\subsubsection{Adversarial Generation of Target Phrases with Different Lengths}

\begin{table}[ht]
\caption{Performance of Adversarial Generation on Target Phrases with Different Length. (Duration results are in Hour:Minute:Second format for all 100 adversarial audio generation. Apart from the duration, the other results are averaged.)}
\label{tab:generation_len}
\centering
\resizebox{.48\textwidth}{!}{
    \begin{tabular}{|p{0.1\textwidth}|p{0.1\textwidth}|p{0.05\textwidth}|p{0.07\textwidth}|p{0.08\textwidth}|}
    \hline
    \textbf{Target Phrase} & \textbf{Success Rate} & \textbf{$\mathbf{dB_{x}(\delta)}$} & \textbf{Duration} & \textbf{Ratio of Frames} \\ \hline
    \multirow{2}{*}{call john smith} & 89.07\% & 29 & 01:10:45 & 100\% \\ \cline{2-5} 
                                     & 91.33\% & 35 & 00:26:09 & 45.65\% \\ \hline
    \multirow{2}{*}{call john} & 84.44\% & 27 & 01:10:25 & 100\% \\ \cline{2-5} 
                               & 85.11\% & 33 & 00:17:42 & 29.54\% \\ \hline
    call john smith & 95.47\% & 32 & 01:11:49 & 100\% \\ \cline{2-5} 
    and david       & 95.52\% & 35 & 00:41:14 & 70.21\% \\ \hline
    \end{tabular}%
}
\end{table}

Some audio adversarial examples are generated corresponding to the selected 100 TIMIT audios with another feature set. To compare the adversarial attack performance on target phrases with different words, we choose the target phase the phrase set named target phrases with different lengths in Table~\ref{tab:target_phrase}. Similar to the above experiment, we evaluate each adversarial example's construction and record their average results in Table~\ref{tab:generation_len}. Assuming that the attacker wants to generate the adversarial example as quickly as possible, the optimization-based method is executed for 1,000 iterations to generate the adversarial example $x'_{begin} = x_{begin} + \delta$. Table \ref{tab:generation_len} lists the performance of FAAG and baseline in the environment with limited computational resources and time.

For the effectiveness analysis, we examine the success rate and distortion results, as listed in Table~\ref{tab:generation_len}. Overall, updating the noise and distortion within 1,000 iterations, all averaged success rates of our method are surpass 85\%, which is comparable to the baseline. The target phrase's different length is not the deterministic factor for either the attack's success rate or the distortion. As discussed in the previous section, the best distortion can be found by adding the $\lambda$ in FAAG for a specific target. In all, no matter how long or short the target phrase is, the performance of our FAAG method is as high as that of the attack proposed in previous work.

For the efficiency analysis, we focus on the duration and $ratio\_frame$ in Table~\ref{tab:generation_len}. Similar to the results in Table~\ref{tab:generation_words}, the ratio of frames used to generate the adversarial examples has a positive correlation to the duration required in the generation. When the target phrase is `call john smith and david', even around 60\% of the whole audio is used for attack, almost half of the time is spent comparing to the baseline. Specifically, different from the baseline, the shorter the target phrase is, the more time is saved using our FAAG method. In all, the FAAG method is less time-consuming than the previous work, especially when the environment is limited to computational resources. 

Over a hundred phrases were collected as the target phrase. These phrases are common voice commands that users usually interact with the surrounding voice assistants, including \textit{Google Assistant} and \textit{Apple Siri} \cite{ok_google, siri_cmd, siri_list}. We randomly selected 100 benign audio clips from the TIMIT dataset and randomly chose each phrase from the collected phrase set for each audio clip as its target phrase. It took 20 minutes and 59 seconds for FAAG to complete all 100 adversarial generations. Compared with the baseline model \cite{carlini2018audio} that spent 50 minutes and 44 seconds, FAAG was much faster in adversarial example generation targeting common command phrases. For effectiveness analysis, FAAG showed a slight advantage on the average success rate ($90.45\% > 88.9\%$) and a slight larger distortion result ($33.11dB > 28.38dB$). In all, the FAAG method is more effective with limited computational resources.

\subsubsection{Comparison with the Baseline using CPUs}
FAAG is an improved method based on the iterative optimization-based method proposed by \cite{carlini2018audio} that is the baseline model in this paper. Empirical results show that our method is significantly more efficient in adversarial audio generation with only one GPU and CPUs. We compare the two methods' performance using CPUs only, primarily focusing on their success rate as $dB_{x}(\delta)$ and time, as listed in Table~\ref{tab:compare_baseline}. Apart from using the CPU only, we compare their performance in a small number of iterations (1,000 iterations) when the target phrase is ``call john smith''. Again, for each adversarial example generation, we repeat ten times and record their averaged results.

\begin{table}[t]
\caption{Comparing the performance of adversarial generation of the audio file \texttt{SA1.wav} with the baseline result. The target phrase is ``call john smith'' for all experiments here. The $Generation \, Time$ is the time for the adversarial audio generation.}
\label{tab:compare_baseline}
\centering
\resizebox{.48\textwidth}{!}{
 \begin{tabular}{|p{0.09\textwidth}|p{0.05\textwidth}|p{0.05\textwidth}|p{0.05\textwidth}|p{0.07\textwidth}|}
 \hline
 \textbf{Generation Method} & \textbf{Ratio Frames} & \textbf{Success Rate} & $\mathbf{dB_{x}(\delta)}$ & \textbf{Generation Time} \\ \hline
 FAAG & 51.96\% & 98.6\% & 28.12 & 26min \\ \hline
 \cite{carlini2018audio} & 100\% & 82.67\% & 29.33 & 70min \\ \hline
 \end{tabular}%
}
\end{table}

Choosing the best $\lambda$ value for the specific audio file \texttt{SA1.wav} with one target phrase, we conduct the results using FAAG. As shown in Table~\ref{tab:compare_baseline}, FAAG significantly exceeds the baseline in \cite{carlini2018audio} from time and success rate aspects using only CPUs with the best $\lambda$. The reason is that FAAG only needs to modify part of the original audio frames, while the baseline needs to modify the whole audio. Without using any GPU, the advance in time using our FAAG is more prominent. As for the success rate and distortion, by choosing the best $\lambda$, FAAG can reach or even surpasses the baseline.

\begin{figure*}[!ht]
  \centering
  \begin{minipage}[b]{0.33\textwidth}
    \includegraphics[width=\textwidth]{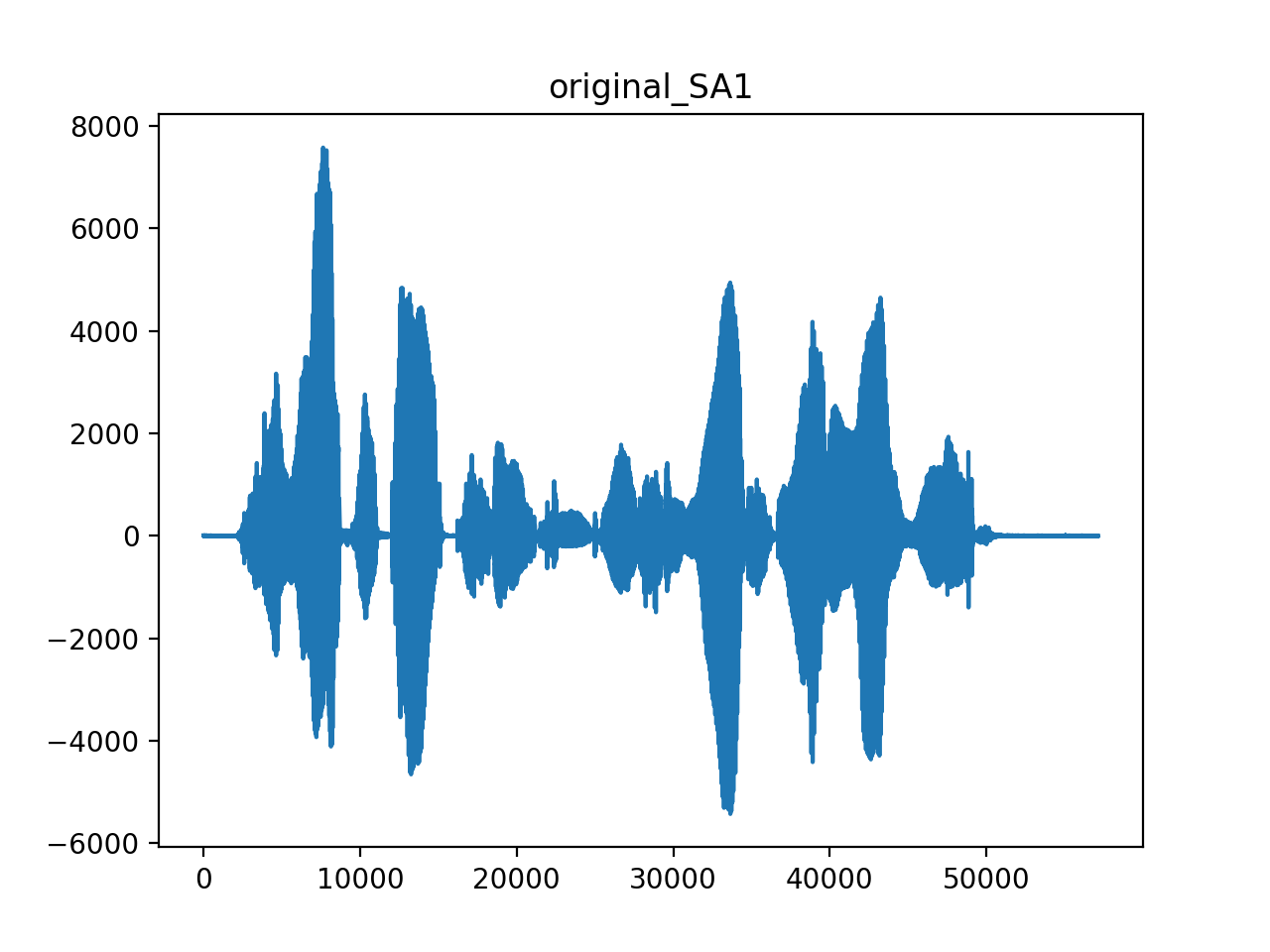}
  \end{minipage}%
  \begin{minipage}[b]{0.33\textwidth}
    \includegraphics[width=\textwidth]{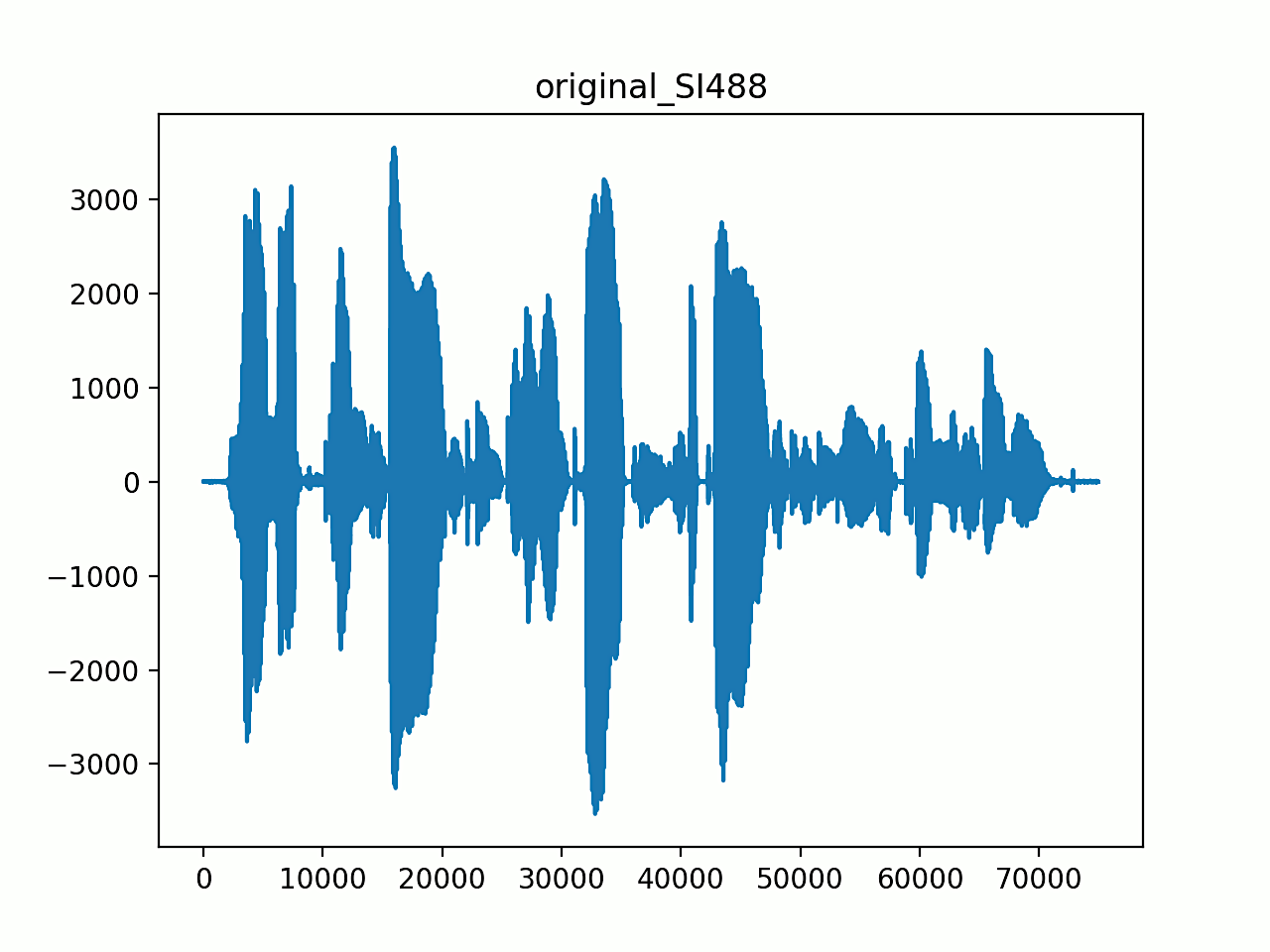}
  \end{minipage}%
  \begin{minipage}[b]{0.33\textwidth}
    \includegraphics[width=\textwidth]{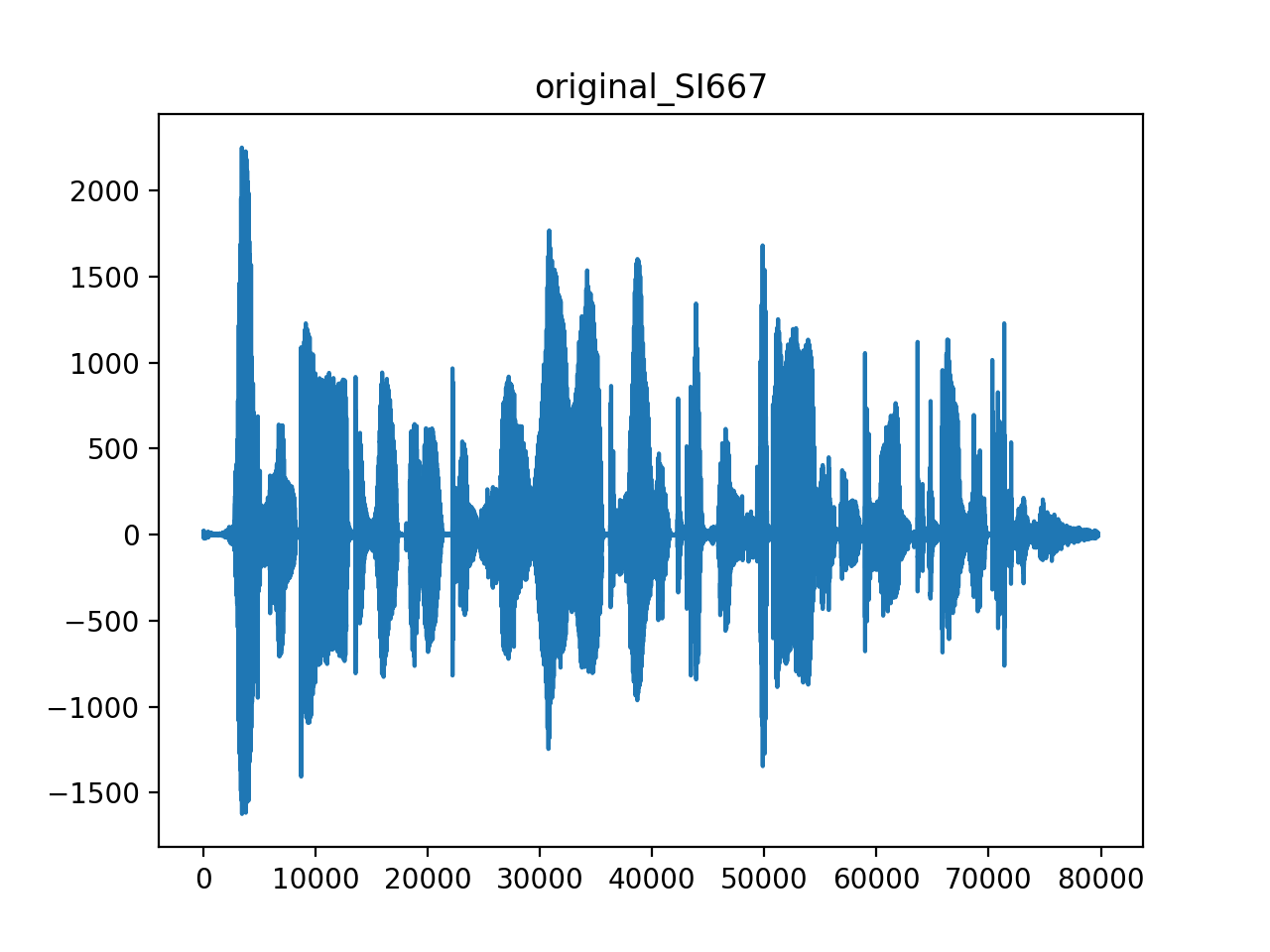}
  \end{minipage}%
  \newline
  \begin{minipage}[b]{0.33\textwidth}
    \includegraphics[width=\textwidth]{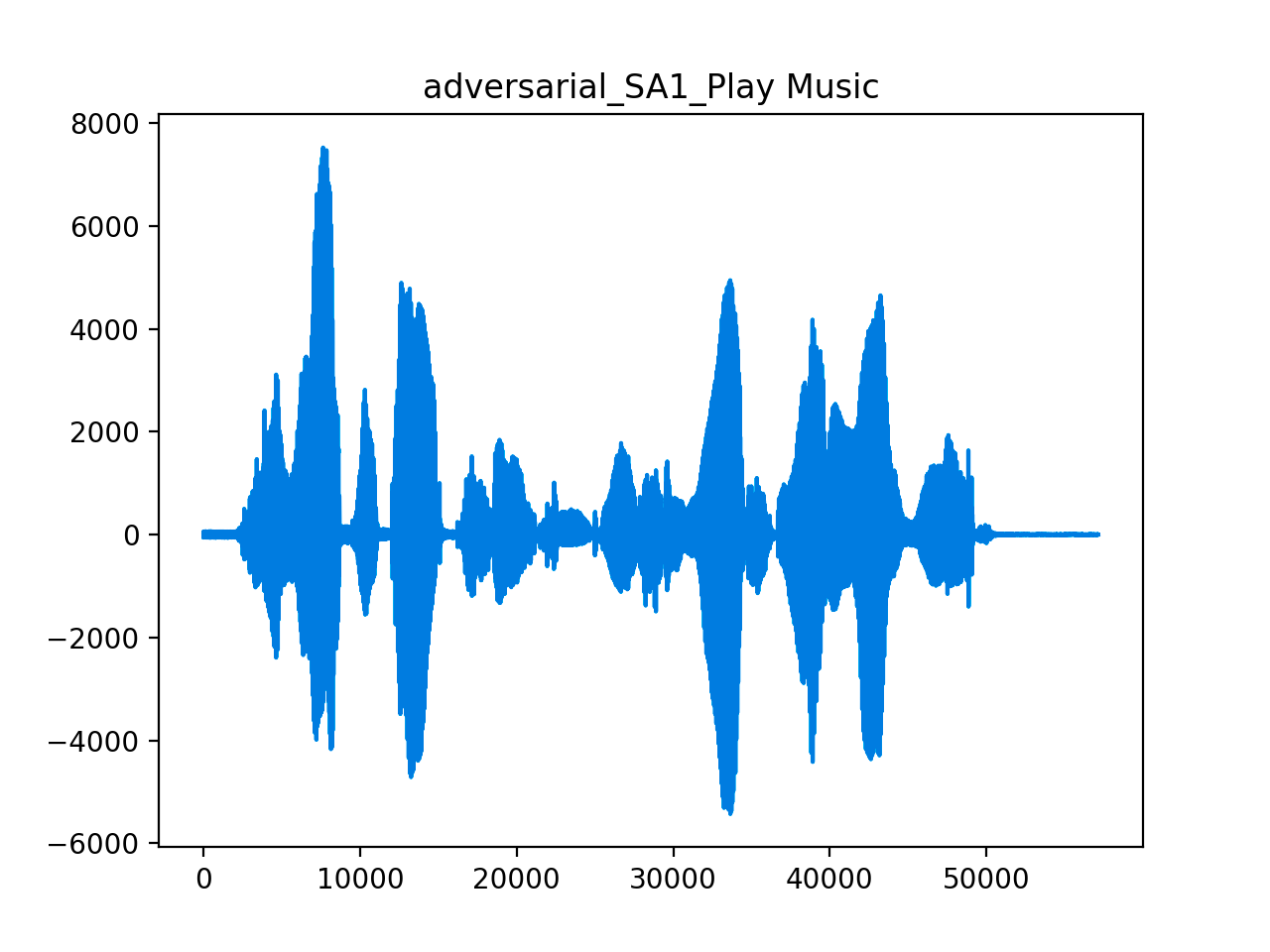}
  \end{minipage}%
  \begin{minipage}[b]{0.33\textwidth}
    \includegraphics[width=\textwidth]{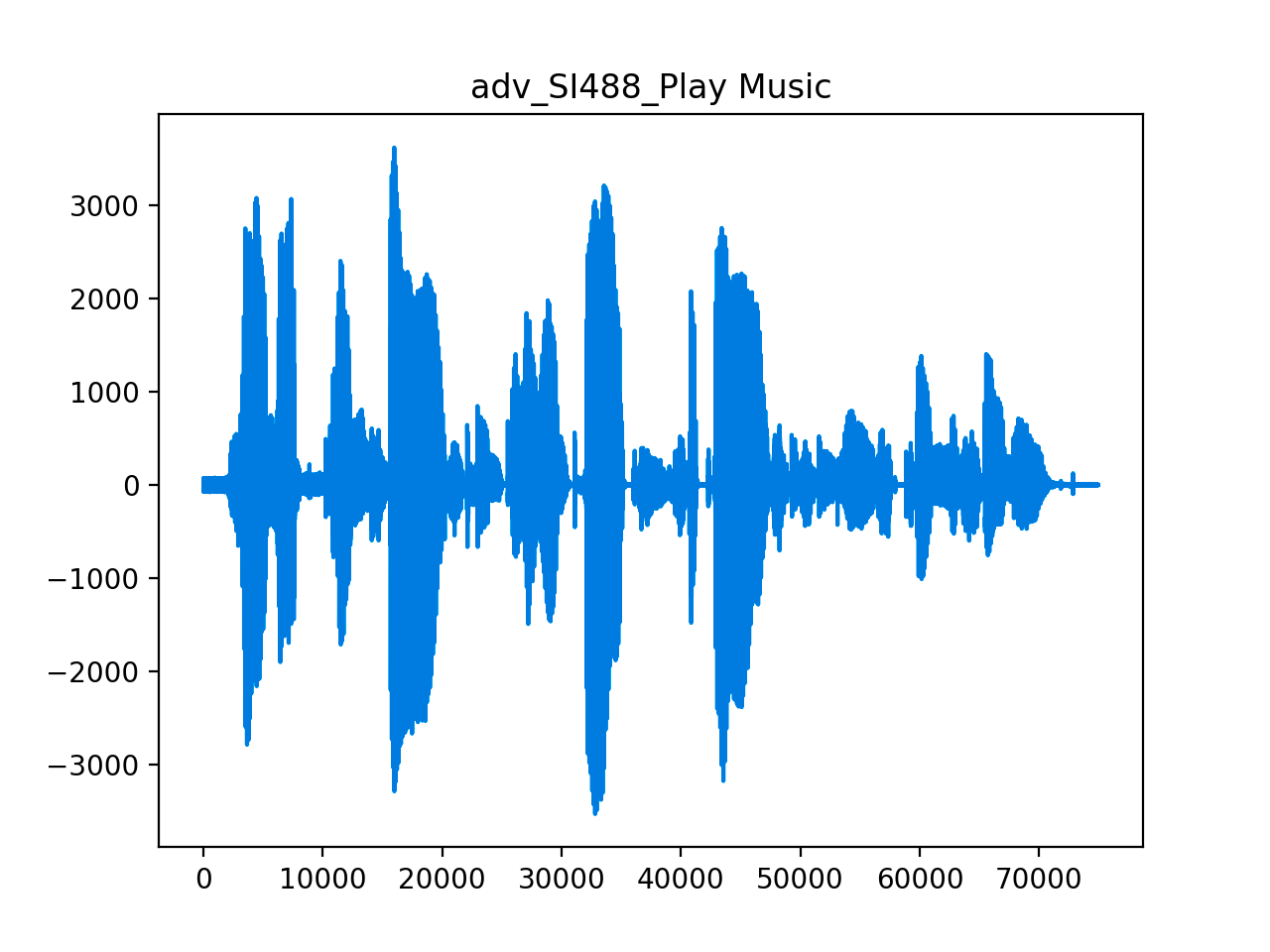}
  \end{minipage}%
  \begin{minipage}[b]{0.33\textwidth}
    \includegraphics[width=\textwidth]{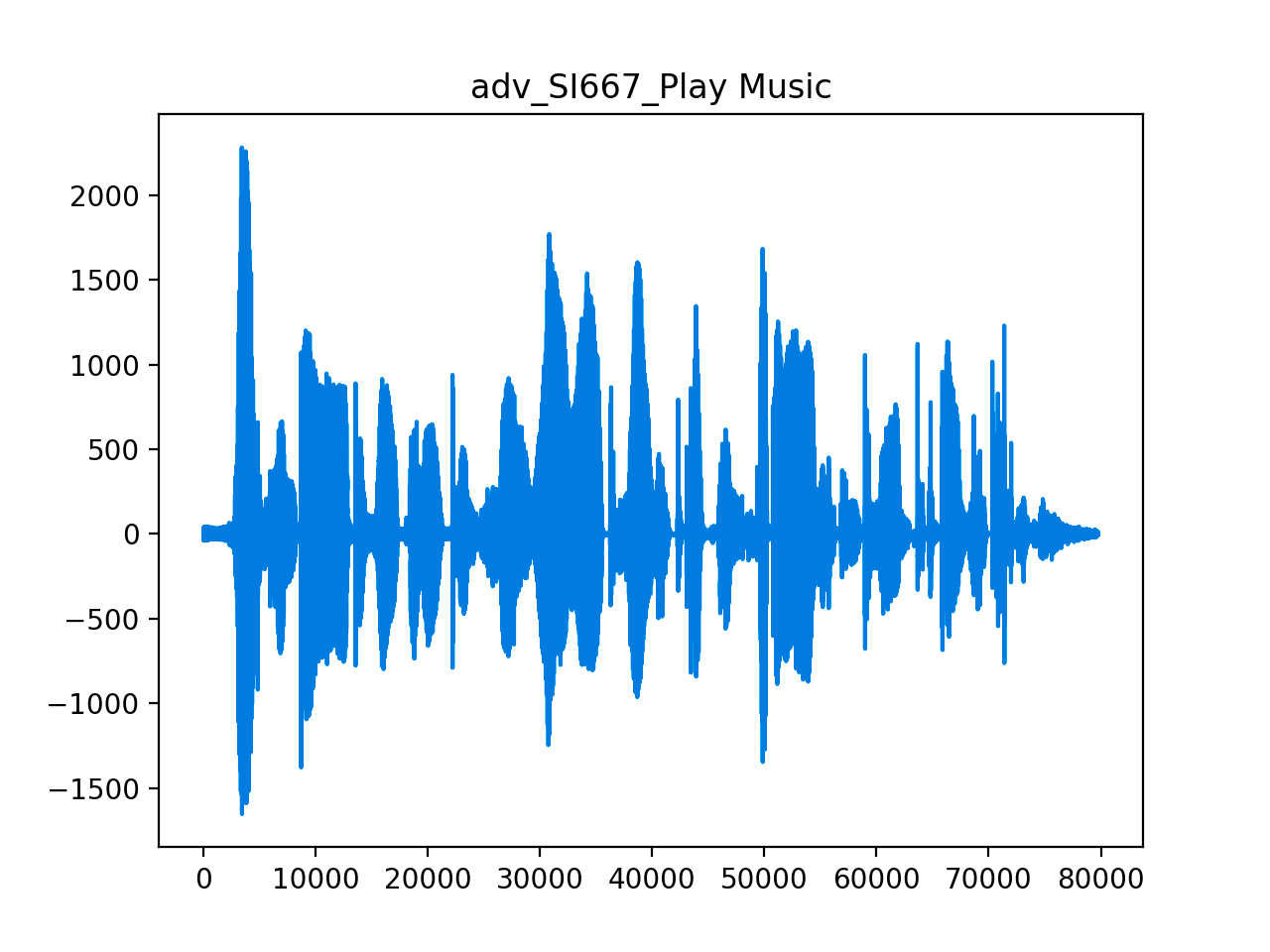}
  \end{minipage}%
  
  \caption{Three pair waves' visualizations. Each column represents a pair of original audio and it adversarial example when the target phrase is ``play music''. The images in the first row present the original waveform of \texttt{SA1.wav}, \texttt{SI488.wav}, and \texttt{SI667.wav}. The images in the second row present the audio waveform of the adversarial audio of corresponding audio.}
  \label{fig:visual}
\end{figure*}

\subsubsection{Case Study}
We present the detailed process of generating adversarial examples with three audio clips \texttt{SA1.wav}, \texttt{SI488.wav}, and \texttt{SI667.wav} with the target phrase ``play music'' as a case study. Assuming that the attacker only uses CPUs to generate the adversarial example for the targeted attack against an ASR model and generates it within one hour. The target ASR model is a DeepSpeech pre-trained ASR model. For the original audio  \texttt{SA1.wav}, the attacker aims to embed the target phrase ``play music'' at the beginning of the original audio by adding noise that will be recognized by the target model but inaudible to human beings.  Algorithm~\ref{alg:select_audio} is used to select a proper clip at the beginning of the original audio $x$ which is marked as $x_{begin} = x - x_{rest}$. Then following the process in Algorithm~\ref{alg:adversarial_generation}, we minimize the problem using the Adam optimizer and the CTC loss function with 100 learning rates and 1,000 iterations. Finally, one adversarial example of the audio \texttt{SA1.wav} is generated. The averaged result is obtained by repeating the previous step ten times.

We randomly select one adversarial example for each \texttt{SA1.wav}, \texttt{SI488.wav} and \texttt{SI667.wav} with the target phrase as ``play music''. Their success rate reached 100\%, 100\%, and 90.9\%. Then we plot the waveform of these adversarial examples comparing to the waveform of their original audio, as shown in Fig.~\ref{fig:visual}. In general, it is challenging to differentiate these two waveforms when the distortion is $34dB$, $36dB$, and $39dB$, respectively. The beginning part of adversarial examples' waveform is slightly thicker than their original audio. Since we only modify the beginning part of the audio, it is faster to generate the adversarial examples than the whole frame's modification.

\subsection{Summary in Speed Advantage}
\skyR{FAAG significantly outperforms the baseline methods in terms of generation time. Table~\ref{tab:compare_time} summarizes the generation time in different conditions using the baseline method \cite{carlini2018audio} and FAAG. When adversarial examples were generated with CPUs only, FAAG can speed up around 56.7\% than the baseline method. When adversarial examples were generated with one GPU and CPUs, FAAG can speed up around 60\% than the baseline method.}

\begin{table}[t]
\caption{Comparison of $Generation\, Time$ between using the baseline method and FAAG. All results are averaged values, while time is in the Hour:Minute:Second format.}
\label{tab:compare_time}
\centering
\resizebox{.48\textwidth}{!}{
 \begin{tabular}{|p{0.09\textwidth}|p{0.13\textwidth}|p{0.05\textwidth}|p{0.1\textwidth}|p{0.09\textwidth}|p{0.06\textwidth}|}
 \hline
 \textbf{Target Dataset} & \textbf{Target Phrase(s)} & \textbf{Using GPU} & \textbf{Generation Time\_Baseline} & \textbf{Generation Time\_FAAG} & \textbf{Speedup} \\ \hline
 \texttt{SA1.wav} & call john smith & Yes & 00:06:03 & 00:02:25 & 60.1\%\\ \hline 
 \texttt{SA1.wav} & call david jone & Yes & 00:06:05 & 00:02:25 & 60.3\% \\ \hline
 \texttt{SA1.wav} & play music list & Yes & 00:06:01 & 00:02:26 & 59.6\% \\ \hline
 \texttt{SI2248.wav} & call john smith & Yes & 00:05:34 & 00:02:24 & 56.9\% \\ \hline
 \texttt{SI667.wav} & call john smith & Yes & 00:08:26 & 00:02:28 & 70.6\% \\ \hline
 \texttt{SA1.wav} & call john smith & No & 01:00:00 & 00:26:00 & 56.7\% \\ \hline
 100 TIMIT audio clips & Phrases with different words & Yes & 01:11:04 & 00:26:20 & 63.4\% \\ \hline
 100 TIMIT audio clips & Phrases with different lengths & Yes & 00:50:44 & 00:20:59 & 60.0\% \\ \hline
 \end{tabular}%
}
\end{table}

\section{Discussion} \label{sec:discussion}

\subsection{Different Position of Adversarial Audio Clip}
Except for hiding the target phrase at the beginning of the audio, we also discuss other audio positions, including the middle part and the ending part. These two positions are also meaningful in some cases. For example, when the target phrase contains a trigger word of an ASR system, the ASR system will only listen to the sentence behind that trigger word. In this section, we present and analyze the results of the last two hiding positions. The original audio file is \texttt{SA1.wav}, and the phrase is `call john smith'.

\begin{figure}[th]
\centering
\includegraphics[width=0.48\textwidth]{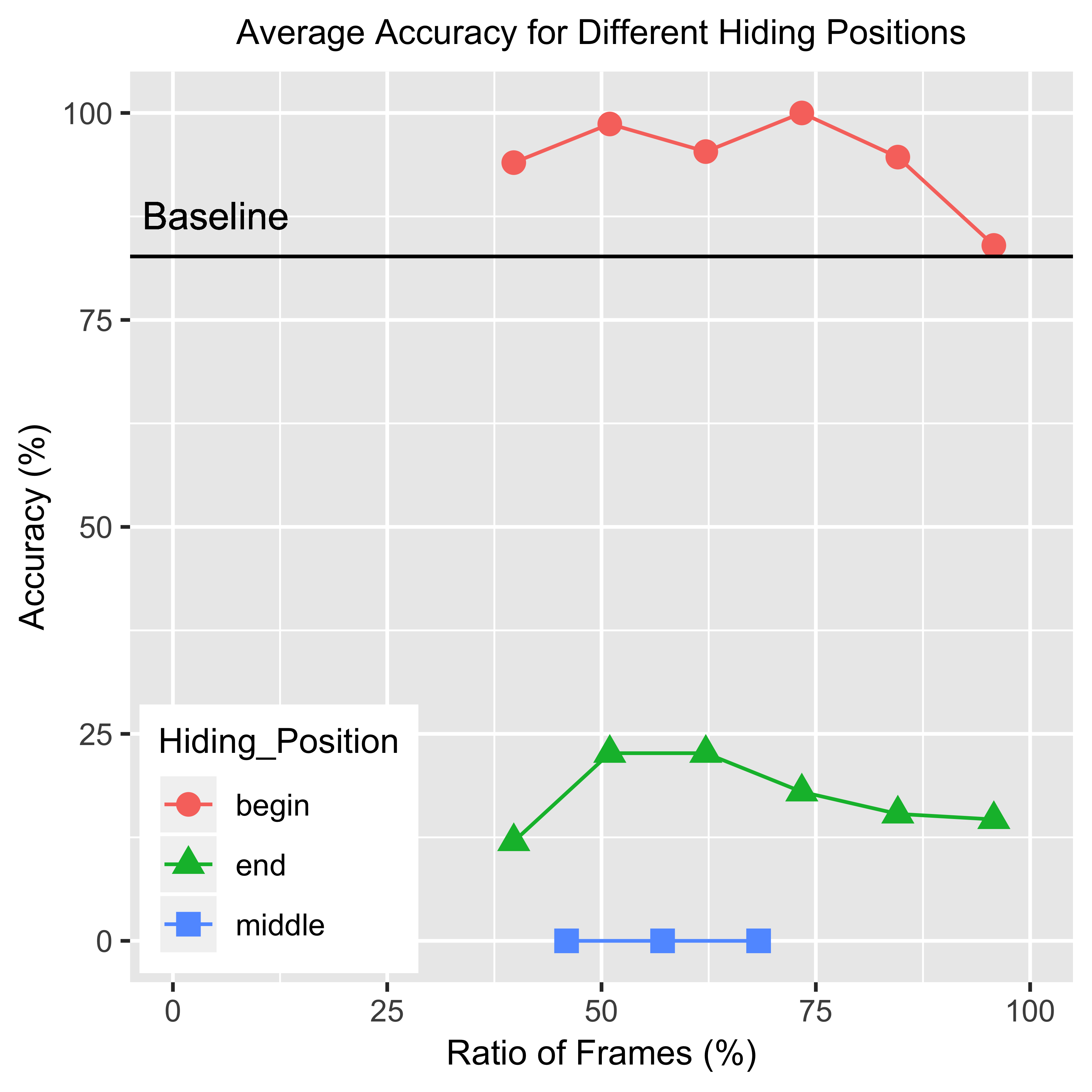}
\caption{Averaged accuracy of the FAAG used in different hiding positions. The baseline is the result using the entire audio sample to generate the adversarial audio \cite{carlini2018audio}.}
\label{fig:positions}
\end{figure}

The selection of audio frames clipped from the middle and the end of the audio is slightly different from the beginning one. For hiding the phrase at the end of the audio, we append two spaces at the beginning of the phrase as the final target phrase. The frame length of audio clips selection is quite similar to the Algorithm~\ref{alg:select_audio}. The differences are $index = |\boldsymbol{x}|-|\boldsymbol{x_{end}}$ and $\boldsymbol{x_{end}} \gets \boldsymbol{x}[index:]$. For hiding the phrase in the middle of the audio, we append two spaces at the beginning and the end of the phrase, respectively, as the final target phrase. The frame length determination is similar to the one hiding in the beginning position. The final index calculates in the Algorithm~\ref{alg:select_audio} are totally different. Another index is introduced for the middle position generation, marked as $index'$. Assume corresponding three characters of output logits will not be replaced. Then $index'=\frac{|\boldsymbol{c}|}{|\boldsymbol{y}|} \times 3 \times s$ and $index=index + \frac{|\boldsymbol{c}|}{|\boldsymbol{y}|} \times |\boldsymbol{t_{allocated}}| \times s$. Two other audio clips and one target audio clip are separated from the original audio, marking as $\boldsymbol{x_{rest}}, \boldsymbol{x_{middle}}, \boldsymbol{x_{rest'}}$. Specifically, $\boldsymbol{x_{rest}} \gets \boldsymbol{x}[:index']$, $\boldsymbol{x_{middle}} \gets \boldsymbol{x}[index':index]$, and $\boldsymbol{x_{rest'} \gets \boldsymbol{x}[index:]}$.

As shown in Fig.~\ref{fig:positions}, only hiding the phrase from the beginning of the audio reaches or surpasses the baseline result. The success rates of hiding the phrase at the end of the audio are always lower than 25\%. Specifically, the first-word `call' cannot be recognized in most cases. Under this situation, the attack is meaningless because the ASR system listing the sentences after the trigger word. As for hiding the phrase in the middle of the audio, no matter how long the frame length is clipped for generation, the adversarial attack did not work. We infer that this is due to the powerful language model within the end-to-end ASR system. Considering the contextual information, especially the previous text, the latter part of the transcription will be influenced.

\subsection{Countermeasures}
Because of the dramatic decrease in the attack's success rate, we find an effective protection method against the current targeted adversarial attack against an end-to-end ASR system. We can append benign audio at the beginning of this suspicious audio before playing it for any suspicious audio. Although the original transcription accuracy will be influenced, the success rate of the targeted attack decreases.

To evaluate the protection results, we append benign audio to the generated adversarial examples and classify them with the target ASR model. Specifically, we select clean audio named \texttt{SA1.wav} and append its complete audio frames to the beginning of any suspicious audios. If the suspicious audio is benign, we expect that the appending audio would not decrease the benign audio's transcription result. Otherwise, if the suspicious audio is an adversarial example, we expect that the combined audio transcription would not include the attacker's target phrase. Assume the attacker's target phrase is ``call john smith''. The suspicious audios include benign audio, several adversarial audios generated using the baseline method \cite{carlini2018audio}, and several adversarial audios generated using our method FAAG. Each experiment repeats ten times, and the averaged results are presented.

\begin{table}[t]
\caption{The performance of various audio files are compared. \texttt{SA1.wav} is the benign audio that we selected to append to the suspicious audios. \texttt{SA1.wav+SA1.wav} represents two begin audios' combination. $Phrase$ represents the target phrase, while $Trans$ is the final translated text. $Success\ Rate$ represents the attack's success rate, while the $Accuracy\ in\ Trans$ comparing the predicted translation with the true translation.}
\label{tab:protect}
\centering
\resizebox{.48\textwidth}{!}{
 \begin{tabular}{|p{0.15\textwidth}|p{0.1\textwidth}|p{0.1 \textwidth}|p{0.1\textwidth}|}
 \hline
 \textbf{Audios} & \textbf{Phrase in Trans} & \textbf{Success Rate} & \textbf{Accuracy in Trans} \\ \hline
 \texttt{SA1.wav} & No & None & 98.04\% \\ \hline
 \texttt{SA1.wav+SA1.wav} & No & None & 96.12\% \\ \hline
 \texttt{Baseline} & Yes & 82.67\% & None  \\ \hline
 \texttt{SA1.wav+Baseline} & No & None & 92.03\% \\ \hline
 \texttt{FAAG} & Yes & 94.45\% & None  \\ \hline
 \texttt{SA1.wav+FAAG} & No & None & 92.52\% \\ \hline
 \end{tabular}%
}
\end{table}

Table~\ref{tab:protect} shows the performance of various audios comparing their translations to the target phrase and their true translation text. In general, appending a complete sample of benign audio at the beginning of any suspicious audio is an effective countermeasure against the targeted adversarial attack. Specifically, comparing the results of \texttt{SA1.wav} and \texttt{SA1.wav+SA1.wav}, this protection method has only a slight decline in translation accuracy. Comparing \texttt{Baseline} with \texttt{SA1.wav+Baseline} and \texttt{FAAG} with \texttt{SA1.wav+Baseline}, the target phrase is not included in the final translation by appending the begin audio. Meanwhile, by appending benign audio, the accuracy of the original audio's translation surpasses 90\%.

\subsection{Transferable FAAG}
FAAG aims to achieve a fast adversarial audio generation. According to Section~\ref{sec:method}, the proper length selection associates with the target model's structure, the benign audio's true transcription, and the target phrase. This paper investigates a recurrent network-based ASR model which is targeted by an iterative optimization-based method. To investigate FAAG's transferability, we adopt another similar iterative optimization-based attack against the DeepSpeech model. Further, we will discuss how to find the proper frame length targeting generic ASR models, like a conventional ASR model.

Yakura and Sakuma \cite{akura2019robust} proposed a robust adversarial audio generation method by extending the iterative optimization-based method \cite{carlini2018audio}. The attack was evaluated in an over-the-air setup. The reported success rate reached 100\% at the cost of more than 18 hours of computational time for each audio. Various techniques like band-pass filters, impulse response, and white Gaussian noise were applied in \cite{akura2019robust} to achieve the best outcomes for very few audio clips. To compare with Yakura and Sakuma's method, we chose a small set of audio clips with one target phrase.

\begin{table}[t]
\caption{Investigating the performance of FAAG and Yakura and Sakuma's method \cite{akura2019robust}. The target phrase is ``call nine one one'' selected from the common voice commands used in the previous experiments. Ten benign audio clips were randomly selected from the TIMIT dataset. All the reported results are averaged.}
\label{tab:compare_physical}
\centering
\resizebox{.48\textwidth}{!}{
 \begin{tabular}{|p{0.09\textwidth}|p{0.05\textwidth}|p{0.05\textwidth}|p{0.05\textwidth}|p{0.07\textwidth}|}
 \hline
 \textbf{Generation Method} & \textbf{Ratio Frames} & \textbf{Success Rate} & $\mathbf{dB_{x}(\delta)}$ & \textbf{Generation Time} \\ \hline
 \cite{akura2019robust} & 100\% & 72.06\% & 84.56 & 30min \\ \hline
 \cite{akura2019robust}+FAAG & 63.69\% & 70.89\% & 79.94 & 21min \\ \hline
 \end{tabular}%
}
\end{table}

We selected one phrase from the common voice commands used in the previous experiment, i.e., ``call nine one one". Ten benign audio clips were randomly selected from the TIMIT dataset. The target ASR model is \texttt{deepspeech-0.4.1}, different from \cite{akura2019robust}. Yakura and Sakuma's method requires adjusting perturbation magnitudes if either the input sample or the target phase changes. Thus, we also adjusted perturbation magnitudes for the audio clips chosen for FAAG. This experiment was designed to measure the speed of the generation method, so over-the-air attack was not evaluated. To ensure fair and reproducible comparisons by avoiding the influences of environmental noises, we used the direct translating method, which is different from the over-the-air evaluation in \cite{akura2019robust}. Moreover, the same number of training iterations were applied for generating each adversarial audio example. Twenty adversarial audio examples were generated for each audio file by using Yakura and Sakuma's method. Feeding these adversarial examples directly into the ASR model, the best example was selected by comparing their transcriptions with the target phrase. Then, FAAG was used with the same setup. The averaged results of the ten audio clips are shown in Table~\ref{tab:compare_physical}. The results demonstrated that FAAG was faster than Yakura and Sakuma's method while generating adversarial examples of similar quality.

When the target model is changed to a different structure, the selected frame length selection will differ. For instance, Kaldi \cite{povey2011kaldi} is a conventional ASR model based on hidden Markov models (HMMs). Since there are no logit outputs to the CTC decoder, Equation \ref{eq:x'_range} no longer holds for Kaldi. Therefore, a different scheme is required to find an explicit relationship among phoneme, HMM-state, pdf identifier, the true transcription, the benign audio, and the target phrase. We leave this investigation as our future work.

\section{Conclusion and Future Work} \label{sec:conclude}

We propose a novel method named FAAG to generate adversarial examples for audio clips through white-box access. In particular, we propose a novel algorithm to determine the appropriate length of the audio frame for adversarial attacks, according to the target phrase, the original audio, and the logit outputs of the target ASR model. The adversarial example can reach a high success rate by adding negligible noise, and the generation process can be completed within a short period using only CPUs with one GPU or even without GPUs. Our empirical studies proved that adding noise to part of the audio can effectively generate an adversarial example. Meanwhile, the distortion caused by the generated adversarial examples is similar to the baseline method, implying little quality loss. Plus, FAAG maintains the adversarial example generation's success rate despite the choice of words in target phrases. More importantly, different positions of hiding phrases are discussed. Hiding the phrase at the beginning of the audio is a plausible attack. Accordingly, appending benign audio to the beginning of adversarial audio can effectively protect service from targeted adversarial audio attack. We verify the adversarial example generated over-the-line to ensure the correctness of the results. Work on conducting the attack over-the-air and under black-box access will be left to the future.



\bibliographystyle{IEEEtran}
\bibliography{main}

\section*{Appendices}

\subsection*{A. The DeepSpeech Model's Architecture}
\sky{The target ASR model used in this paper is Baidu's pre-trained DeepSpeech model named \texttt{deepspeech-0.4.1-model}~\cite{hannun2014deep}. The deepspeech-0.4.1-model is trained by a large Recurrent Neural Network (RNN) using four training corpus, including Fisher, LibriSpeech, Switchboard Hub5'00 , and the English Common Voice. The RNN model is composed of 5 layers of hidden units. Except for the fourth layer, which is a bi-directional recurrent layer, the first three layers and the last layer are not recurrent. The first three layers use a clipped rectified-linear (ReLu) activation function, while the output layer uses a standard softmax function. During the training process, the CTC loss function is used to measure the prediction loss, while Nesterov’s Accelerated gradient method is used for optimization. The dropout rate is 0.15 used in the first three layers. The learning rate is 0.0001, while the training epochs are 30 \cite{deepspeech-model}. As for the performance, this model achieves an 8.26\% word error rate when testing with the LibriSpeech clean test corpus \cite{deepspeech-model}.}

\end{document}